\documentclass[onecolumn]{aa-press}
\usepackage{amssymb}
\usepackage{amsmath}
\usepackage{graphicx,times,natbib}

\input{tcilatex}
\begin{document}

\title{Master equation theory applied to the redistribution of polarized radiation in the weak radiation field limit}

\subtitle{V. The two-term atom}

\author{V\'eronique Bommier}

\institute{LESIA, Observatoire de Paris, PSL Research University, CNRS, Sorbonne Universit\'es, UPMC Univ. Paris 06, Univ. Paris Diderot, Sorbonne Paris Cit\'e 
\newline 5, Place Jules Janssen, 92190 Meudon, France}

\titlerunning{Redistribution of polarized radiation V. 
The two-term atom}
\authorrunning{V. Bommier}

\date{Received ...; accepted ...}

\abstract
{In previous papers of this series, we presented a formalism able to account for both statistical equilibrium of a multilevel atom and coherent and incoherent scatterings (partial redistribution).}
{This paper provides theoretical expressions of the redistribution function for the two-term atom. This redistribution function includes both coherent ($R_{\mathrm{II}}$) and incoherent ($R_{\mathrm{III}}$) scattering contributions with their branching ratios.}
{The expressions were derived by applying the formalism outlined above. The statistical equilibrium equation for the atomic density matrix is first formally solved in the case of the two-term atom with unpolarized and infinitely sharp lower levels. Then the redistribution function is derived by substituting this solution for the expression of  the emissivity.}
{Expressions are provided for both magnetic and non-magnetic cases. Atomic fine structure is taken into account. Expressions are also separately provided under zero and non-zero hyperfine structure.}
{Redistribution functions are widely used in radiative transfer codes. In our formulation, collisional transitions between Zeeman sublevels within an atomic level (depolarizing collisions effect) are taken into account when possible (i.e., in the non-magnetic case). However, the need for a formal solution of the statistical equilibrium as a preliminary step prevents us from taking into account collisional transfers between the levels of the upper term. Accounting for these collisional transfers could be done via a numerical solution of the statistical equilibrium equation system.}


\offprints{V. Bommier, \email{V.Bommier@obspm.fr}}

\maketitle

\section{Introduction}

The redistribution function in radiative transfer expresses the probability
of appearance of an emitted or scattered photon of frequency $\nu $ and
direction $\vec{\Omega}$ in the presence of an incoming photon of frequency $%
\nu _{1}$ and direction $\vec{\Omega}_{1}$. The incoming and the outgoing
photons can be polarized. In this work, we follow the convention of
describing polarized light through the Stokes vector $%
(S_{0},S_{1},S_{2},S_{3})\equiv (I,Q,U,V)$, where $I$\ is the radiation
intensity, $Q$\ and $U$\ are two linearly independent states of linear
polarization, and $V$\ represents circular polarization.

The first expressions of the redistribution function in different cases were
heuristically derived by \citet{Hummer-62}, without taking into account
polarization. He distinguished frequency coherent scattering, denoted as
case II with corresponding redistribution function $R_{\mathrm{II}}$ which
includes a $\delta (\nu -\nu _{1})$ frequency conservation function, and
incoherent scattering, denoted as case III with redistribution function $R_{%
\mathrm{III}}$, where the frequencies of the incoming and outgoing photons
are independent.\ \citet{Omont-etal-72} gave a quantum mechanical basis to
this formulation, and they were able to provide the branching ratios that
weight both coherent and incoherent scattering contributions (see their
well-known Eq. (60) for the case of an infinitely sharp lower level). The
general case where both frequency coherent and incoherent scattering
contribute is known as partial frequency redistribution (PRD).\ When only
incoherent scattering contributes, which is the case of scattering in
spectral line cores, it is known as complete frequency redistribution (CRD).

The redistribution function governs the scattering term of the radiative
transfer equation \citep[see, e.g.,][]{Mihalas-78}. As a consequence,
redistribution functions are widely used in the modeling of radiation from
stellar atmospheres out of local thermodynamical equilibrium (NLTE).\
However, the mechanism modeled by the redistribution function is scattering,
where a photon is absorbed and then reemitted in a spectral line, possibly
coherently in the far wings. The description of the two coupled processes
(absorption and emission) requires the solution of the statistical
equilibrium of the atomic levels as a preliminary step. For a two-level
atom, there is formally only one statistical equilibrium equation expressing
the atomic density matrix of the upper level in terms of that of the lower
level. Its analytic solution is readily derived (even in the presence of
polarization), and the redistribution function follows as a result. In the
case of a greater number of levels, the solution becomes much more involved.
Attempts were done for the three-level atom. An analytic solution is still
possible in this case. This is the equivalent two-level approximation %
\citep{Hubeny-etal-83}. For more than three levels the solution can only be
achieved numerically.

The case of the two-term atom with unpolarized and infinitely sharp lower
levels is a particular case of multilevel atom.\ Each level is defined by
its quantum numbers $(L,S,J)$ with a given orbital quantum number for each
term, $L_{\ell }$ in the lower term and $L_{u}$ in the upper term. $S$ is
the spin angular momentum quantum number, which for allowed transitions is
the same for both terms; and $J$ is the total angular momentum quantum
number defined by $\vec{J}=\vec{L}+\vec{S}$. Due to this vector addition of
angular momenta, there can be multiple values of $J$\ for both the upper ($%
J_{u},J_{u}^{\prime },\ldots $) and lower ($J_{\ell },J_{\ell }^{\prime
},\ldots $) terms. This structure is responsible for $D_{1}$-$D_{2}$ pairs
of lines with $L_{\ell }=0,L_{u}=1$ and $S=1/2$, which results in one single
lower level $J_{\ell }=1/2$ and two upper levels $J_{u}=1/2$ and $%
J_{u}^{\prime }=3/2$,\ as for the \ion{Na}{i} $D_{1}$-$D_{2}$ lines. Other
atoms or ions can be seen in such a structure, and their interest was
recently increased by the observations by %
\citet{Stenflo-etal-00b,Stenflo-etal-00a}, who recorded some linear
polarization in the so-called second solar spectrum of some $D_{1}$ lines,
which are in principle unpolarizable having $J_{u}=1/2<1$. The second solar
spectrum is the spectrum of the linear polarization formed by scattering and
observed close to the solar limb, as defined in \citet{Stenflo-Keller-97}.

The interest raised by this problem led several teams to investigate the
question by modeling the linear polarization formed by scattering in such
lines. However, traditional methods of radiative transfer with partial
redistribution based on the works of \citet{Hummer-62} and 
\citet[Eq.
(60)]{Omont-etal-72} cannot be readily adapted to the treatment of polarized
line formation in these complex two-term transitions, which need to take
into account the presence of fine structure as well as the effect of the
magnetic field.\ Empirical attempts at generalizing the redistribution
function to the polarized case, also including the effects of fine structure
and magnetic fields %
\citep{Smitha-etal-11a,Smitha-etal-13a,Belluzzi-TrujilloB-14} have been
proposed. Hyperfine structure was also considered %
\citep{Smitha-etal-12b,Sowmya-etal-14a}. These generalizations were
exploited in a series of papers for comparison with the observations %
\citep{Smitha-etal-11b,Smitha-etal-12a,Belluzzi-etal-12,Belluzzi-TrujilloB-12,Smitha-etal-13b,Belluzzi-TrujilloB-13,Smitha-etal-14,Sowmya-etal-14b,Belluzzi-etal-15}%
.

However, when compared with the results of a formal theoretical derivation,
it appears that these empirical expressions are not fully correct. The aim
of the present paper is to publish the correct expressions derived from
first principles. The correction concerns essentially the branching ratios
that weight the frequency coherent $R_{\mathrm{II}}$ and incoherent $R_{%
\mathrm{III}}$ terms.\ In fact, we find a contribution of the magnetic field
(Zeeman effect)\ and of the fine (or hyperfine) structure inside these
branching ratios, which were missed by \citet{Smitha-etal-13a} and partly
missed by \citet{Belluzzi-TrujilloB-14}.

\citet{AlsinaB-etal-16} recently studied the polarization of the Mg II k
line, which is a $D_{2}$-type line, including the effects of a magnetic
field. Because their study was spectrally close to the line, they applied
the two-level formalism of \citet{Bommier-97b}. \citet{delPinoA-etal-16}
studied the polarization of the full Mg II h-k doublet, also taking into
account the effects of fine structure and quantum interference in the upper
term. They apply the formalism of \citet{Casini-etal-14}, who derived a
theoretical redistribution function including polarization, fine and
hyperfine structure. However, the final result of \citet{Casini-etal-14} can
be applied only in the collisionless regime (pure $R_{\mathrm{II}}$). %
\citet{delPinoA-etal-16} made use of a generalization of the formalism of %
\citet{Casini-etal-14} to include collisional effects by modifying the
branching ratios in a \textquotedblleft physically
consistent\textquotedblright\ manner (following their words at the top of
the left column, p. 2), but without providing explicit forms of these
branching ratios. Providing these ratios is the object of the present paper.
Recently, \citet{Casini-etal-17} reinvestigated this question in greater
detail. Their final result is given in a compact form by their Eq. (20),
which is the same as Eq. (1) of \citet{delPinoA-etal-16}. It is comprised of
three different added contributions, as is our final result (see Eq. (\ref%
{eq -- redist fine}) and seq.), which is, however, presented here in its
fully explicit algebraic form.

The object of the present paper is to derive branching ratios and associated
redistribution functions from the multilevel/multiline general formalism of
statistical equilibrium of the atomic density matrix and radiative transfer
equations recently published by \citet{Bommier-16a}, derived itself from
first principles following the method outlined in \citet{Bommier-97a}. For
the derivation of the redistribution function, a necessary first step is to
formally solve the statistical equilibrium problem for the atomic density
matrix of the upper state. To this end, the lower state is assumed to be
unpolarized and infinitely sharp (Sect. \ref{section -- stateq});
polarization and magnetic fields are considered there. Section \ref{section
-- with mag field} provides the redistribution function in the presence of a
magnetic field. Section \ref{section -- zero mag field} provides instead the
redistribution function in the particular case of zero magnetic field. In
the concluding Sect. \ref{section -- conclusion}, we outline the main
limitation of this formalism, which is that collisional transfer between the
two fine-structure upper levels cannot be taken into account. In the solar
atmosphere, these collisional transfers are due to collisions with neutral
hydrogen atoms, which are also responsible for level depolarization due to
collisional transitions between the Zeeman sublevels inside a given level.
The two rates for the depolarizing collisions in a given level and the
collisional transfer between the upper term fine-structure levels are,
unfortunately, of the same order of magnitude 
\citep[see, e.g.,][Eqs.
(11-13)]{Kerkeni-Bommier-02}. The effects of level depolarizing collision
rates can be taken into account in the non-magnetic case (which is why the
results of Section \ref{section -- zero mag field} are not trivially an
extension of the magnetic case of Section \ref{section -- with mag field}).
However, collisional transfers cannot be taken into account at all because
their presence prevents the derivation of a formal solution of the
statistical equilibrium system of equations\ because several upper levels
are coupled by these collisional transfers.\ A numerical solution of the
statistical equilibrium system of equations is required for the solution of
the full problem where all required rates are considered. A numerical
solution of the statistical equilibrium equations coupled with the radiative
transfer equation is under way \citep{Bommier-16b}. This new numerical
approach to the solution of the coupled problem does not make use of a
redistribution function.

It should be noted that our present derivation is free of any flat spectrum
approximation about the incident radiation field.

\section{Two-term statistical equilibrium analytical resolution}

\label{section -- stateq}

\subsection{Expression of the statistical equilibrium equation}

We write the statistical equilibrium equation for an excited-level coherence 
$^{\alpha _{u}J_{u}J_{u}^{\prime }}\rho _{M_{u}M_{u}^{\prime }}$ where $%
\alpha _{u}$ represents the set of quantum numbers necessary to specify the
atomic term configuration, including $L_{u}S$. The lower term is assumed to
be infinitely sharp, i.e. its lifetime is infinite. In other words, the
absorption probability $B(\alpha _{\ell }L_{\ell }\rightarrow \alpha
_{u}L_{u})I_{\nu }$ and the inelastic collision excitation probability $%
C(\alpha _{\ell }L_{\ell }\rightarrow \alpha _{u}L_{u})$ are very small with
respect to the radiative de-excitation probability $A(\alpha
_{u}L_{u}\rightarrow \alpha _{\ell }L_{\ell })$. This assumes that the
radiation field in the medium is weak 
\citep [see the definition in
Sect. 2 of][]{Bommier-97a}.\ Because of the stated condition of the infinite
lifetime of the lower term, we can assume that the lower levels are
completely unpolarized, since the presence of depolarizing collisions will
effectively destroy any atomic polarization in the lower term. Thus, only
diagonal elements $^{\alpha _{\ell }J_{\ell }J_{\ell }}\rho _{M_{\ell
}M_{\ell }}$ of the lower-term atomic density matrix must be considered.
From Eq. (9) of \citet{Bommier-16a} one then has%
\begin{equation}
\begin{array}{l}
\medskip \dfrac{\mathrm{d}}{\mathrm{d}t}\ ^{\alpha _{u}J_{u}J_{u}^{\prime
}}\rho _{M_{u}M_{u}^{\prime }}\left( \vec{r},\vec{v}\right) =-\dfrac{\mathrm{%
i}\Delta E_{M_{u}M_{u}^{\prime }}}{\hbar }\ ^{\alpha _{u}J_{u}J_{u}^{\prime
}}\rho _{M_{u}M_{u}^{\prime }}\left( \vec{r},\vec{v}\right) \\ 
\medskip +\dsum\limits_{J_{\ell }M_{\ell }}^{{}}\ ^{\alpha _{\ell }J_{\ell
}J_{\ell }}\rho _{M_{\ell }M_{\ell }}\left( \vec{r},\vec{v}\right) 3B(\alpha
_{\ell }L_{\ell }\rightarrow \alpha _{u}L_{u})(2L_{\ell }+1)(2J_{\ell }+1)%
\sqrt{(2J_{u}+1)(2J_{u}^{\prime }+1)} \\ 
\medskip \times \left\{ 
\begin{array}{ccc}
J_{u} & 1 & J_{\ell } \\ 
L_{\ell } & S & L_{u}%
\end{array}%
\right\} \left\{ 
\begin{array}{ccc}
J_{u}^{\prime } & 1 & J_{\ell } \\ 
L_{\ell } & S & L_{u}%
\end{array}%
\right\} \left( 
\begin{array}{ccc}
J_{u} & 1 & J_{\ell } \\ 
-M_{u} & p & M_{\ell }%
\end{array}%
\right) \left( 
\begin{array}{ccc}
J_{u}^{\prime } & 1 & J_{\ell } \\ 
-M_{u}^{\prime } & p^{\prime } & M_{\ell }%
\end{array}%
\right) \\ 
\medskip \times (-1)^{J_{u}-J_{u}^{\prime }}\dint \mathrm{d}\nu _{1}\doint 
\dfrac{\mathrm{d}\vec{\Omega}_{1}}{4\pi }\mathcal{I}_{-p-p^{\prime }}(\nu
_{1},\vec{\Omega}_{1})\left[ \dfrac{1}{2}\Phi _{ba}\left( \nu
_{M_{u}^{\prime }M_{\ell }}-\tilde{\nu}_{1}\right) +\dfrac{1}{2}\Phi
_{ba}^{\ast }\left( \nu _{M_{u}M_{\ell }}-\tilde{\nu}_{1}\right) \right] \\ 
\medskip -\dfrac{1}{2}\dsum\limits_{J_{u}^{\prime \prime }}\left\{
\dsum\limits_{J_{\ell }M_{\ell }}\ ^{\alpha _{u}J_{u}^{\prime \prime
}J_{u}^{\prime }}\rho _{M_{u}M_{u}^{\prime }}\left( \vec{r},\vec{v}\right)
A(\alpha _{u}L_{u}\rightarrow \alpha _{\ell }L_{\ell })(2L_{u}+1)(2J_{\ell
}+1)\sqrt{(2J_{u}+1)(2J_{u}^{\prime \prime }+1)}\right. \\ 
\medskip \times (-1)^{J_{u}^{\prime \prime }-J_{u}}\left\{ 
\begin{array}{ccc}
J_{u} & 1 & J_{\ell } \\ 
L_{\ell } & S & L_{u}%
\end{array}%
\right\} \left\{ 
\begin{array}{ccc}
J_{u}^{\prime \prime } & 1 & J_{\ell } \\ 
L_{\ell } & S & L_{u}%
\end{array}%
\right\} \left( 
\begin{array}{ccc}
J_{\ell } & 1 & J_{u} \\ 
-M_{\ell } & -p & M_{u}%
\end{array}%
\right) \left( 
\begin{array}{ccc}
J_{\ell } & 1 & J_{u}^{\prime \prime } \\ 
-M_{\ell } & -p & M_{u}%
\end{array}%
\right) \\ 
\medskip +\dsum\limits_{J_{\ell }M_{\ell }}\ ^{\alpha _{u}J_{u}J_{u}^{\prime
\prime }}\rho _{M_{u}M_{u}^{\prime }}\left( \vec{r},\vec{v}\right) A(\alpha
_{u}L_{u}\rightarrow \alpha _{\ell }L_{\ell })(2L_{u}+1)(2J_{\ell }+1)\sqrt{%
(2J_{u}+1)(2J_{u}^{\prime \prime }+1)} \\ 
\medskip \times \left. (-1)^{J_{u}^{\prime \prime }-J_{u}^{\prime }}\left\{ 
\begin{array}{ccc}
J_{u}^{\prime } & 1 & J_{\ell } \\ 
L_{\ell } & S & L_{u}%
\end{array}%
\right\} \left\{ 
\begin{array}{ccc}
J_{u}^{\prime \prime } & 1 & J_{\ell } \\ 
L_{\ell } & S & L_{u}%
\end{array}%
\right\} \left( 
\begin{array}{ccc}
J_{\ell } & 1 & J_{u}^{\prime } \\ 
-M_{\ell } & -p^{\prime } & M_{u}^{\prime }%
\end{array}%
\right) \left( 
\begin{array}{ccc}
J_{\ell } & 1 & J_{u}^{\prime \prime } \\ 
-M_{\ell } & -p^{\prime } & M_{u}^{\prime }%
\end{array}%
\right) \right\}%
\end{array}%
\ .  \label{eq -- stateq1}
\end{equation}%
The first line of this equation accounts for the oscillation effect due to
the possible presence of a magnetic field (the Hanle effect). Lines 2-4
describe the creation of the upper level atomic coherence under the effect
of radiation absorption from the lower term, and lines 5-8 describe the
coherence relaxation under the effect of spontaneous emission. Collisions
are neglected for the moment. The atomic density matrix element depends on
the atom position $\vec{r}$ and velocity $\vec{v}$.

We denote $\Delta E_{M_{u}M_{u}^{\prime }}$ the energy difference between
the two levels $\alpha _{u}J_{u}M_{u}$ and $\alpha _{u}J_{u}^{\prime
}M_{u}^{\prime }$%
\begin{equation}
\Delta E_{M_{u}M_{u}^{\prime }}=E(\alpha _{u}J_{u}M_{u})-E(\alpha
_{u}J_{u}^{\prime }M_{u}^{\prime })\ ,
\end{equation}%
where 
\begin{equation}
E(\alpha _{u}J_{u}M_{u})=E(\alpha _{u}J_{u};B=0)+h\nu _{\mathrm{L}%
}g_{J_{u}}M_{u}\ ,  \label{eq -- Zeeman}
\end{equation}%
where $E(\alpha _{u}J_{u};B=0)$\ is the fine structure level energy in the
absence of a magnetic field, $\nu _{\mathrm{L}}$\ is the Larmor frequency%
\begin{equation}
\nu _{\mathrm{L}}=\frac{\left\vert e\right\vert B}{2\pi m_{e}}\ ,
\end{equation}%
where $\left\vert e\right\vert $\ is the electron charge absolute value, $%
m_{e}$\ is the electron mass, $B$\ is the magnetic field strength, and $%
g_{J_{u}}$\ is the Land\'{e} factor of the upper level $\alpha _{u}J_{u}$.
In Eq. (\ref{eq -- Zeeman}) above, the incomplete Paschen-Back effect is
ignored; it is addressed below in Sect. \ref{subsect -- Paschen-Back}. This
effect has to be taken into account when the Zeeman splitting and the fine
or hyperfine splitting are of the same order of magnitude.

Similarly, $\nu _{M_{u}M_{\ell }}$ denotes the frequency of the transition
between the two levels $\alpha _{u}J_{u}M_{u}$ and $\alpha _{\ell }J_{\ell
}M_{\ell }$, and $\nu _{M_{u}^{\prime }M_{\ell }}$ that of the transition
between the two levels $\alpha _{u}J_{u}^{\prime }M_{u}^{\prime }$ and $%
\alpha _{\ell }J_{\ell }M_{\ell }$%
\begin{equation}
\left\{ 
\begin{array}{c}
\medskip \nu _{M_{u}M_{\ell }}=\left[ E(\alpha _{u}J_{u}M_{u})-E(\alpha
_{\ell }J_{\ell }M_{\ell })\right] /h \\ 
\nu _{M_{u}^{\prime }M_{\ell }}=\left[ E(\alpha _{u}J_{u}^{\prime
}M_{u}^{\prime })-E(\alpha _{\ell }J_{\ell }M_{\ell })\right] /h%
\end{array}%
\ \right. ,
\end{equation}%
where $h$ is the Planck constant. The expression $I_{-p-p^{\prime }}(\tilde{%
\nu}_{1},\vec{\Omega}_{1})$ denotes the tensor of the incident radiation of
frequency $\nu _{1}$ and direction $\vec{\Omega}_{1}$. This tensor is
defined in Sect. 5.11 of the monograph by \citet{Landi-Landolfi-04}, devoted
to the spherical tensors for polarimetry. It is related to the Stokes
parameters of the incident radiation by%
\begin{equation}
\mathcal{I}_{qq^{\prime }}(\nu _{1},\vec{\Omega}_{1})=\dsum\limits_{j=0}^{3}%
\mathcal{T}_{qq^{\prime }}\left( j,\vec{\Omega}_{1}\right) S_{j}\left( \nu
_{1},\vec{\Omega}_{1}\right) \ ,
\end{equation}%
where $S_{j}\ (j=0,1,2,3)$ is one of the Stokes parameters, and $%
T_{qq^{\prime }}\left( j,\vec{\Omega}_{1}\right) $ is the spherical tensor
for polarimetry defined and tabulated by 
\citet[Table 5.3 p.
206]{Landi-Landolfi-04}. Given the direction $\vec{\Omega}_{1}$ of the
incident radiation, the Doppler effect has to be accounted for in the
absorption profiles by the atom with velocity $\vec{v}$ 
\begin{equation}
\tilde{\nu}_{1}=\nu _{1}\left( 1-\frac{\vec{\Omega}_{1}\cdot \vec{v}}{c}%
\right) \ ,
\end{equation}%
see also Sect. 3.1 of \citet{Bommier-16a} about Doppler (or velocity)
redistribution. The parameter $\Phi _{ba}$ denotes the Lorentz absorption
profile, which has the same width $2\gamma _{ba}$ (full width at half
maximum) for all the components of the same multiplet. The definition of the
width of the profile, which includes the effects of collisions, can be found
in Sect. 3.4 of \citet{Bommier-16a}.

Accounting for%
\begin{equation}
\sum_{M_{\ell }p}(2J_{u}+1)\left( 
\begin{array}{ccc}
J_{\ell } & 1 & J_{u} \\ 
-M_{\ell } & -p & M_{u}%
\end{array}%
\right) \left( 
\begin{array}{ccc}
J_{\ell } & 1 & J_{u}^{\prime \prime } \\ 
-M_{\ell } & -p & M_{u}%
\end{array}%
\right) =\delta _{J_{u}J_{u}^{\prime \prime }}
\end{equation}%
and analogously for $J_{u}^{\prime }M_{u}^{\prime }$, and%
\begin{equation}
\sum_{J_{\ell }}(2L_{u}+1)(2J_{\ell }+1)\left\{ 
\begin{array}{ccc}
J_{u} & 1 & J_{\ell } \\ 
L_{\ell } & S & L_{u}%
\end{array}%
\right\} ^{2}=1\ ,
\end{equation}%
lines 5-8 of the above Eq. (\ref{eq -- stateq1}), which account for the
coherence destruction processes, simply reduce into%
\begin{equation}
-A(\alpha _{u}L_{u}\rightarrow \alpha _{\ell }L_{\ell })\ ^{\alpha
_{u}J_{u}J_{u}^{\prime }}\rho _{M_{u}M_{u}^{\prime }}\left( \vec{r},\vec{v}%
\right) \ ,
\end{equation}%
which is also%
\begin{equation}
-\sum_{J_{\ell }}\frac{1}{2}\left[ A(\alpha _{u}L_{u}J_{u}\rightarrow \alpha
_{\ell }L_{\ell }J_{\ell })+A(\alpha _{u}L_{u}J_{u}^{\prime }\rightarrow
\alpha _{\ell }L_{\ell }J_{\ell })\right] \ ^{\alpha _{u}J_{u}J_{u}^{\prime
}}\rho _{M_{u}M_{u}^{\prime }}\left( \vec{r},\vec{v}\right) \ ,
\label{eq -- As for Js}
\end{equation}%
from%
\begin{equation}
A(\alpha _{u}L_{u}J_{u}\rightarrow \alpha _{\ell }L_{\ell }J_{\ell
})=(2L_{u}+1)(2J_{\ell }+1)\left\{ 
\begin{array}{ccc}
J_{u} & 1 & J_{\ell } \\ 
L_{\ell } & S & L_{u}%
\end{array}%
\right\} ^{2}A(\alpha _{u}L_{u}\rightarrow \alpha _{\ell }L_{\ell })
\label{eq -- AJ->AL}
\end{equation}%
and similarly for $J_{u}^{\prime }$.

The lower level is assumed to be infinitely sharp and unpolarized.
Accordingly, the Zeeman sublevels are equally populated as%
\begin{equation}
^{\alpha _{\ell }J_{\ell }J_{\ell }}\rho _{M_{\ell }M_{\ell }}\left( \vec{r},%
\vec{v}\right) =\frac{1}{(2L_{\ell }+1)(2S+1)}\ .
\label{eq -- lterm population}
\end{equation}%
This also assumes that the lower term remains much more populated than the
upper term. Indeed, the ratio of the upper term population to the lower term
population is at most on the order of the number of photons per mode $\bar{n}
$, and in stellar atmosphere physics this number is usually such that $\bar{n%
}\ll 1$, which is the weak radiation field condition 
\citep[Sect.
2]{Bommier-97a}. As a consequence, the population ratio is $\ll 1$. By
simply neglecting the upper term population with respect to the lower term
one, both the above condition and the normalization condition of the atomic
density matrix $Tr\rho =1$ are satisfied. Outside the weak radiation field
condition, the whole present formalism cannot be applied, as discussed in %
\citet{Bommier-97a}. Moreover, the effect of the approximation is even
weaker when one finally considers the ratios of the Stokes parameters $%
Q/I,U/I,V/I$\ instead of the Stokes parameters themselves. The important
point is that all the lower sublevel populations are equal, and that there
are no off-diagonal elements (coherences) in the lower level density matrix.
Coherences or unequal sublevel populations in the lower level would
invalidate the results of the present paper because they would prevent the
analytical solution of the statistical equilibrium as is done in the
following section. When these elements are non-negligible with respect to
the lower level total population, the statistical equilibrium equations form
a system that has to be numerically solved. Apart from this case, a global
scaling factor could be applied to these lower sublevel populations without
changing the results of the present paper, to which the scaling factor would
also have to be applied.

Another consequence of the large lower term population is that the velocity
distribution in this term remains Maxwellian. Lines 2-4 of the above Eq. (%
\ref{eq -- stateq1}), which account for the coherence creation processes,
can then be rewritten as%
\begin{equation}
\begin{array}{l}
\medskip \dsum\limits_{J_{\ell }M_{\ell }}^{{}}\ 3B(\alpha _{\ell }L_{\ell
}\rightarrow \alpha _{u}L_{u})\dfrac{2J_{\ell }+1}{2S+1}\sqrt{%
(2J_{u}+1)(2J_{u}^{\prime }+1)}\ (-1)^{J_{u}-J_{u}^{\prime }} \\ 
\medskip \times \left\{ 
\begin{array}{ccc}
J_{u} & 1 & J_{\ell } \\ 
L_{\ell } & S & L_{u}%
\end{array}%
\right\} \left\{ 
\begin{array}{ccc}
J_{u}^{\prime } & 1 & J_{\ell } \\ 
L_{\ell } & S & L_{u}%
\end{array}%
\right\} \left( 
\begin{array}{ccc}
J_{u} & 1 & J_{\ell } \\ 
-M_{u} & p & M_{\ell }%
\end{array}%
\right) \left( 
\begin{array}{ccc}
J_{u}^{\prime } & 1 & J_{\ell } \\ 
-M_{u}^{\prime } & p^{\prime } & M_{\ell }%
\end{array}%
\right) \\ 
\medskip \times \dint \mathrm{d}\nu _{1}\doint \dfrac{\mathrm{d}\vec{\Omega}%
_{1}}{4\pi }\dsum\limits_{j=0}^{3}\mathcal{T}_{-p-p^{\prime }}\left( j,\vec{%
\Omega}_{1}\right) S_{j}\left( \nu _{1},\vec{\Omega}_{1}\right) \left[ 
\dfrac{1}{2}\Phi _{ba}\left( \nu _{M_{u}^{\prime }M_{\ell }}-\tilde{\nu}%
_{1}\right) +\dfrac{1}{2}\Phi _{ba}^{\ast }\left( \nu _{M_{u}M_{\ell }}-%
\tilde{\nu}_{1}\right) \right]%
\end{array}%
\ .
\end{equation}

\subsection{Solution of the statistical equilibrium equation}

\subsubsection{In the presence of a magnetic field}

Introducing the spherical tensors for polarimetry 
\citep[see Eq.
(5.156)]{Landi-Landolfi-04}%
\begin{equation}
\mathcal{T}_{-p-p^{\prime }}\left( j,\vec{\Omega}_{1}\right)
=\sum_{KQ}(-1)^{1-p}\frac{\sqrt{2K+1}}{\sqrt{3}}\left( 
\begin{array}{ccc}
1 & 1 & K \\ 
-p & p^{\prime } & Q%
\end{array}%
\right) \mathcal{T}_{-Q}^{K}\left( j,\vec{\Omega}_{1}\right) \ ,
\end{equation}%
the statistical equilibrium equation (\ref{eq -- stateq1}) can be resolved
into%
\begin{equation}
\begin{array}{l}
\medskip \ ^{\alpha _{u}J_{u}J_{u}^{\prime }}\rho _{M_{u}M_{u}^{\prime
}}\left( \vec{r},\vec{v}\right) =\medskip \dfrac{1}{A(\alpha
_{u}L_{u}\rightarrow \alpha _{\ell }L_{\ell })+\dfrac{\mathrm{i}\Delta
E_{M_{u}M_{u}^{\prime }}}{\hbar }} \\ 
\medskip \times \dsum\limits_{J_{\ell }M_{\ell
}}^{{}}\dsum\limits_{K^{\prime }Q}B(\alpha _{\ell }L_{\ell }\rightarrow
\alpha _{u}L_{u})\dfrac{2J_{\ell }+1}{2S+1}\sqrt{(2J_{u}+1)(2J_{u}^{\prime
}+1)}\sqrt{3(2K^{\prime }+1)}\ (-1)^{J_{u}-J_{u}^{\prime }+1-p} \\ 
\medskip \times \left\{ 
\begin{array}{ccc}
J_{u} & 1 & J_{\ell } \\ 
L_{\ell } & S & L_{u}%
\end{array}%
\right\} \left\{ 
\begin{array}{ccc}
J_{u}^{\prime } & 1 & J_{\ell } \\ 
L_{\ell } & S & L_{u}%
\end{array}%
\right\} \left( 
\begin{array}{ccc}
J_{u} & 1 & J_{\ell } \\ 
-M_{u} & p & M_{\ell }%
\end{array}%
\right) \left( 
\begin{array}{ccc}
J_{u}^{\prime } & 1 & J_{\ell } \\ 
-M_{u}^{\prime } & p^{\prime } & M_{\ell }%
\end{array}%
\right) \left( 
\begin{array}{ccc}
1 & 1 & K^{\prime } \\ 
-p & p^{\prime } & Q%
\end{array}%
\right) \\ 
\medskip \times \dint \mathrm{d}\nu _{1}\doint \dfrac{\mathrm{d}\vec{\Omega}%
_{1}}{4\pi }\dsum\limits_{j=0}^{3}\mathcal{T}_{-Q}^{K^{\prime }}(j,\vec{%
\Omega}_{1})S_{j}\left( \nu _{1},\vec{\Omega}_{1}\right) \left[ \dfrac{1}{2}%
\Phi _{ba}\left( \nu _{M_{u}^{\prime }M_{\ell }}-\tilde{\nu}_{1}\right) +%
\dfrac{1}{2}\Phi _{ba}^{\ast }\left( \nu _{M_{u}M_{\ell }}-\tilde{\nu}%
_{1}\right) \right]%
\end{array}%
\ .  \label{eq -- solMM'}
\end{equation}%
Introducing the spherical components of the density matrix $^{\alpha
_{u}J_{u}J_{u}^{\prime }}\rho _{Q}^{K}\left( \vec{r},\vec{v}\right) $ with $%
K=0,1,2,\ldots ,J_{u}+J_{u}^{\prime }$ and $Q=-K,-K+1,\ldots ,K-1,K$ as
defined by \citep[see Eq. (3.97)]{Landi-Landolfi-04}%
\begin{equation}
^{\alpha JJ^{\prime }}\rho _{Q}^{K}=\sum_{MM^{\prime }}(-1)^{J-M}\sqrt{2K+1}%
\left( 
\begin{array}{ccc}
J & J^{\prime } & K \\ 
M & -M^{\prime } & -Q%
\end{array}%
\right) \ ^{\alpha JJ^{\prime }}\rho _{MM^{\prime }}\ ,
\end{equation}%
the spherical component can be written as%
\begin{equation}
\begin{array}{l}
\medskip \ ^{\alpha _{u}J_{u}J_{u}^{\prime }}\rho _{Q}^{K}\left( \vec{r},%
\vec{v}\right) =\dsum\limits_{M_{u}M_{u}^{\prime }}\medskip \dfrac{1}{%
A(\alpha _{u}L_{u}\rightarrow \alpha _{\ell }L_{\ell })+\dfrac{\mathrm{i}%
\Delta E_{M_{u}M_{u}^{\prime }}}{\hbar }} \\ 
\medskip \times \dsum\limits_{J_{\ell }M_{\ell
}}^{{}}\dsum\limits_{K^{\prime }}B(\alpha _{\ell }L_{\ell }\rightarrow
\alpha _{u}L_{u})\dfrac{2J_{\ell }+1}{2S+1}\sqrt{(2J_{u}+1)(2J_{u}^{\prime
}+1)}\sqrt{3(2K+1)(2K^{\prime }+1)}\ (-1)^{J_{u}^{\prime }-M_{u}^{\prime
}+1-p^{\prime }} \\ 
\medskip \times \left\{ 
\begin{array}{ccc}
J_{u} & 1 & J_{\ell } \\ 
L_{\ell } & S & L_{u}%
\end{array}%
\right\} \left\{ 
\begin{array}{ccc}
J_{u}^{\prime } & 1 & J_{\ell } \\ 
L_{\ell } & S & L_{u}%
\end{array}%
\right\} \\ 
\medskip \times \left( 
\begin{array}{ccc}
J_{u} & 1 & J_{\ell } \\ 
-M_{u} & p & M_{\ell }%
\end{array}%
\right) \left( 
\begin{array}{ccc}
J_{u}^{\prime } & 1 & J_{\ell } \\ 
-M_{u}^{\prime } & p^{\prime } & M_{\ell }%
\end{array}%
\right) \left( 
\begin{array}{ccc}
1 & 1 & K^{\prime } \\ 
-p & p^{\prime } & Q%
\end{array}%
\right) \left( 
\begin{array}{ccc}
J_{u} & K & J_{u}^{\prime } \\ 
-M_{u} & Q & M_{u}^{\prime }%
\end{array}%
\right) \\ 
\medskip \times \dint \mathrm{d}\nu _{1}\doint \dfrac{\mathrm{d}\vec{\Omega}%
_{1}}{4\pi }\dsum\limits_{j=0}^{3}\mathcal{T}_{-Q}^{K^{\prime }}(j,\vec{%
\Omega}_{1})S_{j}\left( \nu _{1},\vec{\Omega}_{1}\right) \left[ \dfrac{1}{2}%
\Phi _{ba}\left( \nu _{M_{u}^{\prime }M_{\ell }}-\tilde{\nu}_{1}\right) +%
\dfrac{1}{2}\Phi _{ba}^{\ast }\left( \nu _{M_{u}M_{\ell }}-\tilde{\nu}%
_{1}\right) \right]%
\end{array}%
\ .  \label{eq -- solKQ}
\end{equation}

In the case of the two-level atom, i.e., when there is one single $J_{u}$
and one single $J_{\ell }$, this equation can be reduced in terms of
generalized profiles as defined in \citet{Landi-etal-91}. Equation (37) of
this paper is thus obtained, which is also Eq. (7) of \citet{Bommier-97b}.
This is made possible owing to the following relation, which is valid only
if $\alpha _{u}J_{u}=\alpha _{u}J_{u}^{\prime }$\ 
\begin{equation}
\Delta E_{M_{u}M_{u}^{\prime }}=h\nu _{\mathrm{L}}g_{J_{u}}\left(
M_{u}-M_{u}^{\prime }\right) =h\nu _{\mathrm{L}}g_{J_{u}}Q\ .
\end{equation}%
The factorization in $Q$ of $\Delta E_{M_{u}M_{u}^{\prime }}$ permits
restricting the summation over $M_{u}$\ and $M_{u}^{\prime }$\ to just the
second addendum of Eq. (\ref{eq -- solKQ}), leading to the generalized
profile and also making possible the introduction of the depolarizing
collision factors $D^{(K)}$, as obtained in \citet{Bommier-97b}.\ In the
general case of fine structure, when there are several $\alpha _{u}J_{u}$
levels, it is not possible to factorize $\Delta E_{M_{u}M_{u}^{\prime }}$ in
terms of $Q$ when $\alpha _{u}J_{u}\neq \alpha _{u}J_{u}^{\prime }$ because
the form of $\Delta E_{M_{u}M_{u}^{\prime }}$ is more involved.\ However,
when the magnetic field is zero, $\Delta E_{M_{u}M_{u}^{\prime }}$ does not
depend on the magnetic quantum numbers $M_{u}$ and it simply coincides with $%
\Delta E_{J_{u}J_{u}^{\prime }}$, which again makes it possible to introduce
the depolarizing collision factors $D^{(K)}$, as is done in the following
section. However, the depolarizing effect of collisions is thus only partly
taken into account (see Sect. \ref{section -- conclusion}), and this can be
done only in the absence of a magnetic field. The outcome is the numerical
solution of the statistical equilibrium equations.

\subsubsection{In zero magnetic field}

In the absence of a magnetic field, the $M$ indexes in $\Delta E$ and $\nu $
can be replaced by the $J$ indexes and the sums over the $M$ indexes can be
performed. Then, Eq. (\ref{eq -- solMM'}) reduces to%
\begin{equation}
\begin{array}{l}
\medskip \ ^{\alpha _{u}J_{u}J_{u}^{\prime }}\rho _{M_{u}M_{u}^{\prime
}}\left( \vec{r},\vec{v}\right) =\medskip \dfrac{1}{A(\alpha
_{u}L_{u}\rightarrow \alpha _{\ell }L_{\ell })+\dfrac{\mathrm{i}\Delta
E_{J_{u}J_{u}^{\prime }}}{\hbar }} \\ 
\medskip \times \dsum\limits_{J_{\ell }}^{{}}\dsum\limits_{KQ}B(\alpha
_{\ell }L_{\ell }\rightarrow \alpha _{u}L_{u})\dfrac{2J_{\ell }+1}{2S+1}%
\sqrt{(2J_{u}+1)(2J_{u}^{\prime }+1)}\sqrt{3(2K^{\prime }+1)}\
(-1)^{1+J_{\ell }+2J_{u}-M_{u}^{\prime }} \\ 
\medskip \times \left\{ 
\begin{array}{ccc}
J_{u} & 1 & J_{\ell } \\ 
L_{\ell } & S & L_{u}%
\end{array}%
\right\} \left\{ 
\begin{array}{ccc}
J_{u}^{\prime } & 1 & J_{\ell } \\ 
L_{\ell } & S & L_{u}%
\end{array}%
\right\} \left\{ 
\begin{array}{ccc}
K^{\prime } & J_{u} & J_{u}^{\prime } \\ 
J_{\ell } & 1 & 1%
\end{array}%
\right\} \left( 
\begin{array}{ccc}
J_{u} & K^{\prime } & J_{u}^{\prime } \\ 
-M_{u} & Q & M_{u}^{\prime }%
\end{array}%
\right) \\ 
\medskip \times \dint \mathrm{d}\nu _{1}\doint \dfrac{\mathrm{d}\vec{\Omega}%
_{1}}{4\pi }\dsum\limits_{j=0}^{3}\mathcal{T}_{-Q}^{K^{\prime }}(j,\vec{%
\Omega}_{1})S_{j}\left( \nu _{1},\vec{\Omega}_{1}\right) \left[ \dfrac{1}{2}%
\Phi _{ba}\left( \nu _{J_{u}^{\prime }J_{\ell }}-\tilde{\nu}_{1}\right) +%
\dfrac{1}{2}\Phi _{ba}^{\ast }\left( \nu _{J_{u}J_{\ell }}-\tilde{\nu}%
_{1}\right) \right]%
\end{array}%
\ ,
\end{equation}%
where only the sum over $M_{\ell }$ has been performed, whereas Eq. (\ref{eq
-- solKQ}) becomes%
\begin{equation}
\begin{array}{l}
\medskip \ ^{\alpha _{u}J_{u}J_{u}^{\prime }}\rho _{Q}^{K}\left( \vec{r},%
\vec{v}\right) =\medskip \dfrac{1}{A(\alpha _{u}L_{u}\rightarrow \alpha
_{\ell }L_{\ell })+\dfrac{1}{2}\left[ D^{(K)}(\alpha
_{u}J_{u})+D^{(K)}(\alpha _{u}J_{u}^{\prime })\right] +\dfrac{\mathrm{i}%
\Delta E_{J_{u}J_{u}^{\prime }}}{\hbar }} \\ 
\medskip \times B(\alpha _{\ell }L_{\ell }\rightarrow \alpha _{u}L_{u})%
\dfrac{2J_{\ell }+1}{2S+1}\sqrt{3(2J_{u}+1)(2J_{u}^{\prime }+1)}\
(-1)^{1+J_{\ell }+J_{u}+Q} \\ 
\medskip \times \left\{ 
\begin{array}{ccc}
J_{u} & 1 & J_{\ell } \\ 
L_{\ell } & S & L_{u}%
\end{array}%
\right\} \left\{ 
\begin{array}{ccc}
J_{u}^{\prime } & 1 & J_{\ell } \\ 
L_{\ell } & S & L_{u}%
\end{array}%
\right\} \left\{ 
\begin{array}{ccc}
K & J_{u} & J_{u}^{\prime } \\ 
J_{\ell } & 1 & 1%
\end{array}%
\right\} \\ 
\medskip \times \dint \mathrm{d}\nu _{1}\doint \dfrac{\mathrm{d}\vec{\Omega}%
_{1}}{4\pi }\dsum\limits_{j=0}^{3}\mathcal{T}_{-Q}^{K}(j,\vec{\Omega}%
_{1})S_{j}\left( \nu _{1},\vec{\Omega}_{1}\right) \left[ \dfrac{1}{2}\Phi
_{ba}\left( \nu _{J_{u}^{\prime }J_{\ell }}-\tilde{\nu}_{1}\right) +\dfrac{1%
}{2}\Phi _{ba}^{\ast }\left( \nu _{J_{u}J_{\ell }}-\tilde{\nu}_{1}\right) %
\right]%
\end{array}%
\ ,
\end{equation}%
where the sums over $M_{\ell },M_{u},M_{u}^{\prime }$ have been performed,
and where one has applied%
\begin{equation}
\dsum\limits_{MM^{\prime }}\left( 
\begin{array}{ccc}
J & J^{\prime } & K \\ 
M & -M^{\prime } & -Q%
\end{array}%
\right) \left( 
\begin{array}{ccc}
J & J^{\prime } & K^{\prime } \\ 
M & -M^{\prime } & -Q%
\end{array}%
\right) =\frac{1}{2K+1}\delta _{KK^{\prime }}\ .
\end{equation}

The depolarizing collisions have been taken into account here, by adding
their contribution in the $K$-order coherence inverse lifetime contribution
and similarly to Eq. (\ref{eq -- As for Js}). This is possible only for the
tensorial component $^{\alpha _{u}J_{u}J_{u}^{\prime }}\rho _{Q}^{K}$ but
not for the dyadic component $^{\alpha _{u}J_{u}J_{u}^{\prime }}\rho
_{M_{u}M_{u}^{\prime }}$, as explained below. In addition, this does not
account for all the elastic or quasi-elastic collision effects. In the solar
atmosphere, these effects are due to collisions with neutral hydrogen atoms.
These collisions induce both transitions between Zeeman sublevels and
between fine or hyperfine structure levels. Here we only account for the
transitions between the Zeeman sublevels and we ignore the other
transitions. However, as detailed in Sect. \ref{section -- conclusion}, both
contributions are of the same order of magnitude. Thus, by so doing we
include only a part of the effect of the elastic or quasi-elastic collisions.

An attempt to account for the depolarizing collisions was heuristically
proposed by \citet{Smitha-etal-13a}, but introducing the same $D^{(K)}$ rate
for both $\alpha _{u}J_{u}$ and $\alpha _{u}J_{u}^{\prime }$ states (with $%
D^{(0)}=0$). The above expression is the result of a formal derivation and
accounts for the fine structure as far as possible, i.e., in the absence of
a magnetic field and in tensorial components. This is the only case where
the effect of the transitions between the Zeeman sublevels can be simply
accounted for through the $D^{(K)}(\alpha _{u}J_{u})$ rates. In the general
case, this is not possible due to the presence of $\Delta
E_{M_{u}M_{u}^{\prime }}$ because the explicit dependence on $M_{u}$ and $%
M_{u}^{\prime }$ excessively complicates the formalism. In addition, the
transitions between the fine structure levels cannot be accounted for in any
case. This was pointed out by \citet{Belluzzi-TrujilloB-14} in their Sect.
3.1, and was the reason why they discarded the whole depolarizing effect of
collisions from their formalism. As pointed out by these authors, taking
into account all the effects of the collisions with neutral hydrogen
requires the numerical solution of the statistical equilibrium.

In the general case of fine structure, as stated in the above, it is not
possible to factorize $\Delta E_{M_{u}M_{u}^{\prime }}$ in terms of $Q$ when 
$\alpha _{u}J_{u}\neq \alpha _{u}J_{u}^{\prime }$ and in non-zero magnetic
fields. The statistical equilibrium equations are then simple only in the
dyadic basis. In this basis, the introduction of the depolarizing collisions
effect, even limited to those responsible for the $D^{(K)}$ coefficients,
leads to the coupling of all the upper level Zeeman sublevel coherences, as
visible in the unnumbered equation before Eq. (7.99) of %
\citet{Landi-Landolfi-04}. This coupling prevents an analytical solution of
the statistical equilibrium in the dyadic basis from being attained.
Transforming into the irreducible tensors basis leads to other couplings,
also preventing an analytical solution, due to the non-factorization of $%
\Delta E_{M_{u}M_{u}^{\prime }}$ in terms of $Q$ in the presence of a fine
structure. This is why it is possible to simply introduce $D^{(K)}$
coefficients only in the zero magnetic field case, when there is a fine
structure. The outcome of this problem is the numerical solution of the
statistical equilibrium equations.

\section{Redistribution function in the presence of a magnetic field}

\label{section -- with mag field}

The expression of the redistribution function follows by porting the
solution of the statistical equilibrium in the emissivity, which is given in
Eqs. (15-16) of \citet{Bommier-16a}. The emissivity represents the emitted
light quantity per unit volume and is given by%
\begin{equation}
\varepsilon _{i}\left( \nu ,\vec{\Omega}\right) =k_{L}\dint \mathrm{d}\nu
_{1}\doint \dfrac{\mathrm{d}\vec{\Omega}_{1}}{4\pi }\dsum\limits_{j=0}^{3}%
\mathcal{R}_{ij}\left( \nu ,\nu _{1},\vec{\Omega},\vec{\Omega}_{1};\vec{B}%
\right) S_{j}\left( \nu _{1},\vec{\Omega}_{1}\right) \ ,
\label{eq -- emissivity}
\end{equation}%
where the redistribution function $R_{ij}\left( \nu ,\nu _{1},\vec{\Omega},%
\vec{\Omega}_{1};\vec{B}\right) $ is the joint probability of observing a
photon of frequency $\nu $ and polarization $i$ emitted in the $\vec{\Omega}$
direction, given a photon of frequency $\nu _{1}$ and polarization $j$
incident along the $\vec{\Omega}_{1}$ direction, in the presence of a
magnetic field $\vec{B}$. The parameter $k_{L}$ is the line absorption
coefficient%
\begin{equation}
k_{L}=\frac{h\nu }{4\pi }\mathcal{N}B(\alpha _{\ell }L_{\ell }\rightarrow
\alpha _{u}L_{u})\ ,
\end{equation}%
where $\mathcal{N}$ is the emitter atom or ion density and $B(\alpha _{\ell
}L_{\ell }\rightarrow \alpha _{u}L_{u})$ is the Einstein coefficient for
absorption in the line. We recall that the atom lower term is here assumed
to be overpopulated, i.e., the entire atomic population is assumed to be in
the ground state as defined in Eq. (\ref{eq -- lterm population}); $\mathcal{%
N}$\ is then also the lower term atom or ion density.

The emission can possibly end in level $(\alpha _{\ell }L_{\ell }SJ_{\ell
}^{\prime }M_{\ell }^{\prime })$ of the lower term, which may be different
from the initial $(\alpha _{\ell }L_{\ell }SJ_{\ell }M_{\ell })$ level.

The contribution of inelastic collisions has to be considered. For
de-excitation, this adds a collisional contribution $C(\alpha
_{u}L_{u}\rightarrow \alpha _{\ell }L_{\ell })$ to the radiative
de-excitation probability $A(\alpha _{u}L_{u}\rightarrow \alpha _{\ell
}L_{\ell })$. For excitation, it adds a contribution term besides the
radiative excitation term in Eq. (\ref{eq -- stateq1}).\ This collisional
excitation term is at the origin of the well-known Planck source term of the
radiative transfer equation, which we will not rewrite here 
\citep[see for
instance Eq. (25) of ][]{Belluzzi-TrujilloB-14}.

To simplify notations, we introduce the usual $\Gamma $ probabilities in the
excited term%
\begin{equation}
\left\{ 
\begin{array}{l}
\medskip \Gamma _{R}=A(\alpha _{u}L_{u}\rightarrow \alpha _{\ell }L_{\ell })
\\ 
\medskip \Gamma _{I}=C(\alpha _{u}L_{u}\rightarrow \alpha _{\ell }L_{\ell })
\\ 
\medskip \Gamma _{E}=2\gamma _{ba}^{(c)}%
\end{array}%
\ \right. ,
\end{equation}%
where $\Gamma _{R}$ is the radiative deexcitation coefficient, $\Gamma _{I}$
is the inelastic collisional deexcitation coefficient, and $\Gamma _{E}$ is
the elastic collision contribution to line broadening as described in %
\citet{Sahal-Bommier-14,Sahal-Bommier-17} The expression of $\gamma
_{ba}^{(c)}$ in terms of the collisional $S$ and $T=1-S$ matrices is also
given in Eqs. (28-29) of \citet{Bommier-16a}.

The redistribution function as defined in Eq. (\ref{eq -- emissivity}) is
normalized to 
\begin{equation}
\dint \mathrm{d}\nu \dint \mathrm{d}\nu _{1}\doint \dfrac{\mathrm{d}\vec{%
\Omega}}{4\pi }\doint \dfrac{\mathrm{d}\vec{\Omega}_{1}}{4\pi }\dint \frac{%
\mathrm{d}\gamma }{2\pi }\dint \frac{\mathrm{d}\gamma _{1}}{2\pi }\mathcal{R}%
_{ij}\left( \nu ,\nu _{1},\vec{\Omega},\vec{\Omega}_{1};\vec{B}\right)
=\delta _{i,0}\delta _{j,0}\frac{\Gamma _{R}}{\Gamma _{R}+\Gamma _{I}}\ ,
\end{equation}%
and not to unity. This is the same for the two-level atom redistribution
function defined in \citet{Bommier-97b}. This is also the same in %
\citet{Omont-etal-72}; the normalization condition is given in their Eq.
(64), which is the same normalization condition as above, and $\gamma $\ is
the angle with the reference axis for the linear polarization definition
(the direction of positive polarization) in the plane perpendicular to the
line of sight \citep[see Fig. 2
in][]{Landi-83}.\ The redistribution function is not normalized to unity
because here it accounts for the total emitted radiation. However, the
photon emission generally results from two contributions, scattering of the
incident photon on the one hand, and collisional excitation followed by
radiative de-excitation on the other.\ This explains why the result of the
scattering contribution is not normalized to unity. However, other authors
prefer to normalize their redistribution function to unity 
\citep[see,
e.g.,][Eq. (2-7)]{Mihalas-78}. In this case, the effect of the competing
collisional excitation is accounted for by the $(1-\epsilon )$\ coefficient,
which factorizes the scattering contribution in the transfer equation (see
Eqs. (2-36), (2-41) and (13-92)). Here, $\epsilon =\Gamma _{I}/\left( \Gamma
_{R}+\Gamma _{I}\right) $. In order to apply the formulae we present in the
following, it is convenient to multiply our redistribution functions by $%
\left( \Gamma _{R}+\Gamma _{I}\right) /\Gamma _{R}$. %
\citet{Belluzzi-TrujilloB-14} apply the same normalization as ours to their
redistribution functions (see their Eqs. (1, 7-12) and following unnumbered
equations). As the results of \citet{Casini-etal-14} can only be applied in
the collisionless regime, the two possible normalizations described above
are identical and are both equal to unity in this regime.

The redistribution function normalization condition is based on the
following normalization property of the irreducible spherical tensors for
polarimetry \citep{Landi-84} 
\begin{equation}
\doint \dfrac{\mathrm{d}\vec{\Omega}}{4\pi }\dint \frac{\mathrm{d}\gamma }{%
2\pi }\mathcal{T}_{Q}^{K}(i,\vec{\Omega})=\delta _{i,0}\delta _{K,0}\delta
_{Q,0}\ ,
\end{equation}%
as can be derived for instance from Table 1 in \citet{Bommier-97b}.

Considering both the second-order $\varepsilon _{i}^{\left( 2\right) }$ and
fourth-order $\varepsilon _{i}^{\left( 4\right) }$ contributions to the
emissivity as defined in Eqs. (15-16) of \citet{Bommier-16a}, the following
expressions can be derived, which contain both frequency coherent $R_{%
\mathrm{II}}$ and frequency incoherent $R_{\mathrm{III}}$ and their
branching ratios.

\subsection{Considering only fine structure}

Let us first provide a few formulae useful for the calculation. In the
fourth-order emissivity, there is a profile $\Phi _{ca}\left( \tilde{\nu}-%
\tilde{\nu}_{1}-\nu _{\alpha _{\ell }J_{\ell }M_{\ell },\alpha _{\ell
}J_{\ell }^{\prime }M_{\ell }^{\prime }}\right) $. In the case of the
two-term atom, $c=a$ and the profile width is the lower term width $\gamma
_{a}$, which tends towards zero because we have assumed that the lower level
is infinitely sharp $\gamma _{a}\longrightarrow 0$. Therefore, 
\begin{equation}
\frac{1}{2}\Phi _{a}\left( \tilde{\nu}-\tilde{\nu}_{1}-\nu _{M_{\ell
}M_{\ell }^{\prime }}\right) =\frac{1}{\gamma _{a}-2\mathrm{i\pi }\left( 
\tilde{\nu}-\tilde{\nu}_{1}-\nu _{M_{\ell }M_{\ell }^{\prime }}\right) }%
\longrightarrow \frac{1}{2}\delta \left( \tilde{\nu}-\tilde{\nu}_{1}-\nu
_{M_{\ell }M_{\ell }^{\prime }}\right) +\frac{\mathrm{i}}{2\pi }\mathcal{P}%
\frac{1}{\tilde{\nu}-\tilde{\nu}_{1}-\nu _{M_{\ell }M_{\ell }^{\prime }}}\ ,
\end{equation}%
where $\mathcal{P}$ stands for the Cauchy principal value.

Equation (3) in \citet{Bommier-97b}, which transforms a profile product into
a profile sum, is also useful for the calculation. In the present notation,
this is%
\begin{equation}
\dfrac{1}{2}\Phi _{ba}\left( \nu _{M_{u}^{\prime }M_{\ell }}-\tilde{\nu}%
_{1}\right) \ \dfrac{1}{2}\Phi _{ba}^{\ast }\left( \nu _{M_{u}M_{\ell }}-%
\tilde{\nu}_{1}\right) =\frac{1}{2\gamma _{ba}+\dfrac{\mathrm{i}\Delta
E_{M_{u}M_{u}^{\prime }}}{\hbar }}\left[ \dfrac{1}{2}\Phi _{ba}\left( \nu
_{M_{u}^{\prime }M_{\ell }}-\tilde{\nu}_{1}\right) +\dfrac{1}{2}\Phi
_{ba}^{\ast }\left( \nu _{M_{u}M_{\ell }}-\tilde{\nu}_{1}\right) \right] \ ,
\end{equation}%
where $2\gamma _{ba}$ is the full line width at half maximum 
\citep[see
Sect. 3.4 of][]{Bommier-16a}, and $2\gamma _{ba}=\Gamma _{R}+\Gamma
_{I}+\Gamma _{E}$.

Considering only the line fine structure, one obtains%
\begin{equation}
\begin{array}{l}
\medskip \mathcal{R}_{ij}\left( \nu ,\nu _{1},\vec{\Omega},\vec{\Omega}_{1};%
\vec{B}\right) =\dsum\limits_{J_{u}M_{u}J_{u}^{\prime }M_{u}^{\prime
}J_{\ell }M_{\ell }J_{\ell }^{\prime }M_{\ell }^{\prime }KK^{\prime }Q}\int
f(\vec{v})\mathrm{d}^{3}\vec{v}\ (-1)^{Q}\mathcal{T}_{-Q}^{K^{\prime }}(j,%
\vec{\Omega}_{1})\mathcal{T}_{Q}^{K}(i,\vec{\Omega}) \\ 
\medskip \times 3\dfrac{2L_{u}+1}{2S+1}(2J_{u}+1)(2J_{u}^{\prime
}+1)(2J_{\ell }+1)(2J_{\ell }^{\prime }+1)\sqrt{(2K+1)(2K^{\prime }+1)}\
(-1)^{M_{\ell }-M_{\ell }^{\prime }} \\ 
\medskip \times \left\{ 
\begin{array}{ccc}
J_{u} & 1 & J_{\ell } \\ 
L_{\ell } & S & L_{u}%
\end{array}%
\right\} \left\{ 
\begin{array}{ccc}
J_{u}^{\prime } & 1 & J_{\ell } \\ 
L_{\ell } & S & L_{u}%
\end{array}%
\right\} \left\{ 
\begin{array}{ccc}
J_{u} & 1 & J_{\ell }^{\prime } \\ 
L_{\ell } & S & L_{u}%
\end{array}%
\right\} \left\{ 
\begin{array}{ccc}
J_{u}^{\prime } & 1 & J_{\ell }^{\prime } \\ 
L_{\ell } & S & L_{u}%
\end{array}%
\right\} \\ 
\medskip \times \left( 
\begin{array}{ccc}
J_{u} & 1 & J_{\ell } \\ 
-M_{u} & p & M_{\ell }%
\end{array}%
\right) \left( 
\begin{array}{ccc}
J_{u}^{\prime } & 1 & J_{\ell } \\ 
-M_{u}^{\prime } & p^{\prime } & M_{\ell }%
\end{array}%
\right) \left( 
\begin{array}{ccc}
J_{u} & 1 & J_{\ell }^{\prime } \\ 
-M_{u} & p^{\prime \prime \prime } & M_{\ell }^{\prime }%
\end{array}%
\right) \left( 
\begin{array}{ccc}
J_{u}^{\prime } & 1 & J_{\ell }^{\prime } \\ 
-M_{u}^{\prime } & p^{\prime \prime } & M_{\ell }^{\prime }%
\end{array}%
\right) \\ 
\medskip \times \left( 
\begin{array}{ccc}
1 & 1 & K^{\prime } \\ 
-p & p^{\prime } & Q%
\end{array}%
\right) \left( 
\begin{array}{ccc}
1 & 1 & K \\ 
-p^{\prime \prime \prime } & p^{\prime \prime } & Q%
\end{array}%
\right) \\ 
\medskip \times \left\{ \dfrac{\Gamma _{R}}{\Gamma _{R}+\Gamma _{I}+\Gamma
_{E}+\dfrac{\mathrm{i}\Delta E_{M_{u}M_{u}^{\prime }}}{\hbar }}\ \delta
\left( \tilde{\nu}-\tilde{\nu}_{1}-\nu _{M_{\ell }M_{\ell }^{\prime
}}\right) \left[ \dfrac{1}{2}\Phi _{ba}\left( \nu _{M_{u}^{\prime }M_{\ell
}}-\tilde{\nu}_{1}\right) +\dfrac{1}{2}\Phi _{ba}^{\ast }\left( \nu
_{M_{u}M_{\ell }}-\tilde{\nu}_{1}\right) \right] \right. \\ 
\medskip +\left[ \dfrac{\Gamma _{R}}{\Gamma _{R}+\Gamma _{I}+\dfrac{\mathrm{i%
}\Delta E_{M_{u}M_{u}^{\prime }}}{\hbar }}-\dfrac{\Gamma _{R}}{\Gamma
_{R}+\Gamma _{I}+\Gamma _{E}+\dfrac{\mathrm{i}\Delta E_{M_{u}M_{u}^{\prime }}%
}{\hbar }}\right] \\ 
\medskip \times \left. \left[ \dfrac{1}{2}\Phi _{ba}\left( \nu
_{M_{u}^{\prime }M_{\ell }}-\tilde{\nu}_{1}\right) +\dfrac{1}{2}\Phi
_{ba}^{\ast }\left( \nu _{M_{u}M_{\ell }}-\tilde{\nu}_{1}\right) \right] %
\left[ \dfrac{1}{2}\Phi _{ba}\left( \nu _{M_{u}^{\prime }M_{\ell }^{\prime
}}-\tilde{\nu}\right) +\dfrac{1}{2}\Phi _{ba}^{\ast }\left( \nu
_{M_{u}M_{\ell }^{\prime }}-\tilde{\nu}\right) \right] \right\}%
\end{array}%
\ .  \label{eq -- redist fine}
\end{equation}

The frequency coherent contribution weighted by its branching ratio is given
by line 6 of this equation, and the frequency incoherent contribution
weighted by its branching ratio is given by lines 7-8. The second branching
ratio is made of two subtracted terms. The first term without $\Gamma _{E}$
in the denominator stems from the second-order contribution to the
emissivity $\varepsilon _{i}^{(2)}$. The second term with $\Gamma _{E}$ in
the denominator, and the frequency coherent contribution of line 6 stem from
the fourth-order contribution to the emissivity $\varepsilon _{i}^{(4)}$.

Similarities can be found between this expression and those proposed by %
\citet{delPinoA-etal-16} and \citet{Casini-etal-17}. They propose a
redistribution function made of three added contributions, as the present
one is. The allocation of the elastic collision rate $\Gamma _{E}$\ in the
branching ratios is similar in the two formalisms. However, one of the
redistribution function contributions proposed by \citet{delPinoA-etal-16}
and \citet{Casini-etal-17} is computed within the flat spectrum
approximation, which means that for this term the incident radiation is
assumed to be free of any spectral lines.\ Our present redistribution
function is free of any flat spectrum approximation. All three contributions
depend on the incident radiation frequency $\tilde{\nu}_{1}.$

It can be seen that the upper term fine structure intervals intervene in the
branching ratios via $\Delta E_{M_{u}M_{u}^{\prime }}$. These fine structure
intervals were completely missed in the heuristic derivation by %
\citet{Smitha-etal-13a}, and partly missed by \citet{Belluzzi-TrujilloB-14},
who also missed some exact atomic frequencies defined in the last line of
the equation (see details below in the zero magnetic field case). The
presence of this term prevents us from carrying out any of the summations
indicated at the beginning of the equation. This prevents any factorization
in the redistribution function. Due to the $M$ indexes the magnetic field
also contributes to this term. When there is no magnetic field, the
summations over the $M$'s can then be performed. This is done in the
following subsection.

\subsection{Considering also hyperfine structure}

The case of non-zero hyperfine structure can easily be derived from the case
of only fine structure described above by making the following
transformations $J\rightarrow F,\ L\rightarrow J,\ S\rightarrow I$, and by
applying Eq. (\ref{eq -- AJ->AL}) above to transform the transition
probabilities into the term transition probability. Thus, we obtain for the
redistribution function%
\begin{equation}
\begin{array}{l}
\medskip \mathcal{R}_{ij}\left( \nu ,\nu _{1},\vec{\Omega},\vec{\Omega}_{1};%
\vec{B}\right) =\dsum\limits_{J_{u}F_{u}M_{u}J_{u}^{\prime }F_{u}^{\prime
}M_{u}^{\prime }J_{\ell }F_{\ell }M_{\ell }J_{\ell }^{\prime }F_{\ell
}^{\prime }M_{\ell }^{\prime }KK^{\prime }Q}\int f(\vec{v})\mathrm{d}^{3}%
\vec{v}\ (-1)^{Q}\mathcal{T}_{-Q}^{K^{\prime }}(j,\vec{\Omega}_{1})\mathcal{T%
}_{Q}^{K}(i,\vec{\Omega}) \\ 
\medskip \times 3\dfrac{2L_{u}+1}{2S+1}(2J_{u}+1)(2J_{u}^{\prime
}+1)(2J_{\ell }+1)(2J_{\ell }^{\prime }+1)(2F_{u}+1)(2F_{u}^{\prime
}+1)(2F_{\ell }+1)(2F_{\ell }^{\prime }+1) \\ 
\medskip \times \sqrt{(2K+1)(2K^{\prime }+1)}\ (-1)^{M_{\ell }-M_{\ell
}^{\prime }} \\ 
\medskip \times \left\{ 
\begin{array}{ccc}
J_{u} & 1 & J_{\ell } \\ 
L_{\ell } & S & L_{u}%
\end{array}%
\right\} \left\{ 
\begin{array}{ccc}
J_{u}^{\prime } & 1 & J_{\ell } \\ 
L_{\ell } & S & L_{u}%
\end{array}%
\right\} \left\{ 
\begin{array}{ccc}
J_{u} & 1 & J_{\ell }^{\prime } \\ 
L_{\ell } & S & L_{u}%
\end{array}%
\right\} \left\{ 
\begin{array}{ccc}
J_{u}^{\prime } & 1 & J_{\ell }^{\prime } \\ 
L_{\ell } & S & L_{u}%
\end{array}%
\right\} \\ 
\medskip \times \left\{ 
\begin{array}{ccc}
F_{u} & 1 & F_{\ell } \\ 
J_{\ell } & S & J_{u}%
\end{array}%
\right\} \left\{ 
\begin{array}{ccc}
F_{u}^{\prime } & 1 & F_{\ell } \\ 
J_{\ell } & S & J_{u}%
\end{array}%
\right\} \left\{ 
\begin{array}{ccc}
F_{u} & 1 & F_{\ell }^{\prime } \\ 
J_{\ell } & S & J_{u}%
\end{array}%
\right\} \left\{ 
\begin{array}{ccc}
F_{u}^{\prime } & 1 & F_{\ell }^{\prime } \\ 
J_{\ell } & S & J_{u}%
\end{array}%
\right\} \\ 
\medskip \times \left( 
\begin{array}{ccc}
F_{u} & 1 & F_{\ell } \\ 
-M_{u} & p & M_{\ell }%
\end{array}%
\right) \left( 
\begin{array}{ccc}
F_{u}^{\prime } & 1 & F_{\ell } \\ 
-M_{u}^{\prime } & p^{\prime } & M_{\ell }%
\end{array}%
\right) \left( 
\begin{array}{ccc}
F_{u} & 1 & F_{\ell }^{\prime } \\ 
-M_{u} & p^{\prime \prime \prime } & M_{\ell }^{\prime }%
\end{array}%
\right) \left( 
\begin{array}{ccc}
F_{u}^{\prime } & 1 & F_{\ell }^{\prime } \\ 
-M_{u}^{\prime } & p^{\prime \prime } & M_{\ell }^{\prime }%
\end{array}%
\right) \\ 
\medskip \times \left( 
\begin{array}{ccc}
1 & 1 & K^{\prime } \\ 
-p & p^{\prime } & Q%
\end{array}%
\right) \left( 
\begin{array}{ccc}
1 & 1 & K \\ 
-p^{\prime \prime \prime } & p^{\prime \prime } & Q%
\end{array}%
\right) \\ 
\medskip \times \left\{ \dfrac{\Gamma _{R}}{\Gamma _{R}+\Gamma _{I}+\Gamma
_{E}+\dfrac{\mathrm{i}\Delta E_{M_{u}M_{u}^{\prime }}}{\hbar }}\ \delta
\left( \tilde{\nu}-\tilde{\nu}_{1}-\nu _{M_{\ell }M_{\ell }^{\prime
}}\right) \left[ \dfrac{1}{2}\Phi _{ba}\left( \nu _{M_{u}^{\prime }M_{\ell
}}-\tilde{\nu}_{1}\right) +\dfrac{1}{2}\Phi _{ba}^{\ast }\left( \nu
_{M_{u}M_{\ell }}-\tilde{\nu}_{1}\right) \right] \right. \\ 
\medskip +\left[ \dfrac{\Gamma _{R}}{\Gamma _{R}+\Gamma _{I}+\dfrac{\mathrm{i%
}\Delta E_{M_{u}M_{u}^{\prime }}}{\hbar }}-\dfrac{\Gamma _{R}}{\Gamma
_{R}+\Gamma _{I}+\Gamma _{E}+\dfrac{\mathrm{i}\Delta E_{M_{u}M_{u}^{\prime }}%
}{\hbar }}\right] \\ 
\medskip \times \left. \left[ \dfrac{1}{2}\Phi _{ba}\left( \nu
_{M_{u}^{\prime }M_{\ell }}-\tilde{\nu}_{1}\right) +\dfrac{1}{2}\Phi
_{ba}^{\ast }\left( \nu _{M_{u}M_{\ell }}-\tilde{\nu}_{1}\right) \right] %
\left[ \dfrac{1}{2}\Phi _{ba}\left( \nu _{M_{u}^{\prime }M_{\ell }^{\prime
}}-\tilde{\nu}\right) +\dfrac{1}{2}\Phi _{ba}^{\ast }\left( \nu
_{M_{u}M_{\ell }^{\prime }}-\tilde{\nu}\right) \right] \right\}%
\end{array}%
\ .  \label{eq -- redist hyperfine}
\end{equation}

In the collisionless regime, this redistribution function is in agreement
with that derived by \citet{Casini-etal-14}, including the branching ratio
(the coefficient before the $\delta $\ function).

\subsection{Transition from the Zeeman effect to the Paschen-Back effect}

\label{subsect -- Paschen-Back}

This is also called the incomplete Paschen-Back effect case. This case was
treated in \citet{Bommier-80}, and we refer to this publication for details.
In this case, the magnetic quantum number $M_{J}$ remains a good quantum
number, but the total angular momentum quantum number $J$ is no longer a
good quantum number. The modified quantum number corresponding to each
Hamiltonian eigentstate $\left\vert \alpha LSJ^{\ast }M_{J}\right\rangle $
can be denoted as $J^{\ast }$ and the eigenstate can be developed over the $%
\left\vert \alpha LSJ^{\prime }M_{J}\right\rangle $ basis states as 
\begin{equation}
\left\vert \alpha LSJ^{\ast }M_{J}\right\rangle =\dsum\limits_{J^{\prime
}}C_{J^{\ast }M_{J}}^{J^{\prime }}\left( B\right) \left\vert \alpha
LSJ^{\prime }M_{J}\right\rangle \ .
\end{equation}%
In other words, $\left\vert \alpha LSJ^{\ast }M_{J}\right\rangle $\ is an
eigenvector of the Hamiltonian, and $C_{J^{\ast }M_{J}}^{J^{\prime }}\left(
B\right) $\ is the expansion coefficient of this eigenvector over the
considered basis vectors $\left\vert \alpha LSJM_{J}\right\rangle $. The
star $\left( \ast \right) $\ qualifies the $J$\ quantum number of the same
level in zero magnetic field. As described in \citet{Bommier-80}, the $%
C_{J^{\ast }M_{J}}^{J^{\prime }}\left( B\right) $ quantity is obtained by
diagonalizing the Hamiltonian of fine structure plus magnetic field
interaction. These Hamiltonians are real quantities and the $C_{J^{\ast
}M_{J}}^{J^{\prime }}\left( B\right) $ coefficients are therefore real. In
the 3-$j$ and 6-$j$ coefficients entering the redistribution function, each $%
J$ has to be replaced by the corresponding $J^{\ast }$, which leads to a
summation of 3-$j$ or 6-$j$ made of the corresponding $J^{\prime }$ and
weighted by the $C_{J^{\ast }M_{J}}^{J^{\prime }}\left( B\right) $
coefficients. This is detailed in Appendix \ref{Appendix}.

A similar effect occurs for the hyperfine structure when the hyperfine
splitting becomes comparable to the magnetic splitting. This case is called
transition from the Zeeman effect to the Back-Goudsmit effect, or incomplete
Back-Goudsmit effect. Similarly,%
\begin{equation}
\left\vert \alpha JIF^{\ast }M\right\rangle =\dsum\limits_{J^{\prime
}}C_{F^{\ast }M}^{F^{\prime }}\left( B\right) \left\vert \alpha JIF^{\prime
}M\right\rangle \ .
\end{equation}

It should be kept in mind that although $J$ and $F$ are no longer good
quantum numbers, in the sense that each eigenvector now has to be decomposed
over the basis vectors, the magnetic quantum numbers $M$ and $M_{J}=M+M_{I}$
remain good quantum numbers, where $M_{I}$\ is the magnetic quantum number
for the nuclear spin and $M_{J}$ is the magnetic quantum number for the
total fine structure angular momentum $J$. Given a state $\left\vert \alpha
LSJ^{\ast }IF^{\ast }M\right\rangle $, it is possible to determine the value
of $M_{J}$. This is detailed in Appendix \ref{Appendix} where we provide the
generalization of the previous Eqs. (\ref{eq -- redist fine}) and (\ref{eq
-- redist hyperfine}) to the case of the incomplete Paschen-Back and
Back-Goudsmit effects.

\section{Redistribution function in zero magnetic field}

\label{section -- zero mag field}

As already stated, in the absence of a magnetic field the $M$ indexes in $%
\Delta E$ and $\nu $ can be replaced by the $J$ indexes and the sums over
all the $M$ indexes can be performed.

\subsection{In considering only fine structure}

In the case of fine structure only and zero magnetic field, we obtain%
\begin{equation}
\begin{array}{l}
\medskip \mathcal{R}_{ij}\left( \nu ,\nu _{1},\vec{\Omega},\vec{\Omega}_{1};%
\vec{B}=\vec{0}\right) =\dsum\limits_{J_{u}J_{u}^{\prime }J_{\ell }J_{\ell
}^{\prime }KQ}\int f(\vec{v})\mathrm{d}^{3}\vec{v}\ (-1)^{Q}\mathcal{T}%
_{-Q}^{K}(j,\vec{\Omega}_{1})\mathcal{T}_{Q}^{K}(i,\vec{\Omega}) \\ 
\medskip \times 3\dfrac{2L_{u}+1}{2S+1}(2J_{u}+1)(2J_{u}^{\prime
}+1)(2J_{\ell }+1)(2J_{\ell }^{\prime }+1)\ (-1)^{J_{\ell }-J_{\ell
}^{\prime }} \\ 
\medskip \times \left\{ 
\begin{array}{ccc}
J_{u} & 1 & J_{\ell } \\ 
L_{\ell } & S & L_{u}%
\end{array}%
\right\} \left\{ 
\begin{array}{ccc}
J_{u}^{\prime } & 1 & J_{\ell } \\ 
L_{\ell } & S & L_{u}%
\end{array}%
\right\} \left\{ 
\begin{array}{ccc}
J_{u} & 1 & J_{\ell }^{\prime } \\ 
L_{\ell } & S & L_{u}%
\end{array}%
\right\} \left\{ 
\begin{array}{ccc}
J_{u}^{\prime } & 1 & J_{\ell }^{\prime } \\ 
L_{\ell } & S & L_{u}%
\end{array}%
\right\} \\ 
\medskip \times \left\{ 
\begin{array}{ccc}
K & J_{u} & J_{u}^{\prime } \\ 
J_{\ell } & 1 & 1%
\end{array}%
\right\} \left\{ 
\begin{array}{ccc}
K & J_{u} & J_{u}^{\prime } \\ 
J_{\ell }^{\prime } & 1 & 1%
\end{array}%
\right\} \\ 
\medskip \times \left\{ \dfrac{\Gamma _{R}}{\Gamma _{R}+\Gamma _{I}+\Gamma
_{E}+\dfrac{\mathrm{i}\Delta E_{J_{u}J_{u}^{\prime }}}{\hbar }}\ \delta
\left( \tilde{\nu}-\tilde{\nu}_{1}-\nu _{J_{\ell }J_{\ell }^{\prime
}}\right) \left[ \dfrac{1}{2}\Phi _{ba}\left( \nu _{J_{u}^{\prime }J_{\ell
}}-\tilde{\nu}_{1}\right) +\dfrac{1}{2}\Phi _{ba}^{\ast }\left( \nu
_{J_{u}J_{\ell }}-\tilde{\nu}_{1}\right) \right] \right. \\ 
\medskip +\left[ \dfrac{\Gamma _{R}}{\Gamma _{R}+\Gamma _{I}+\dfrac{1}{2}%
\left[ D^{(K)}(\alpha _{u}J_{u})+D^{(K)}(\alpha _{u}J_{u}^{\prime })\right] +%
\dfrac{\mathrm{i}\Delta E_{J_{u}J_{u}^{\prime }}}{\hbar }}-\dfrac{\Gamma _{R}%
}{\Gamma _{R}+\Gamma _{I}+\Gamma _{E}+\dfrac{\mathrm{i}\Delta
E_{J_{u}J_{u}^{\prime }}}{\hbar }}\right] \\ 
\medskip \times \left. \left[ \dfrac{1}{2}\Phi _{ba}\left( \nu
_{J_{u}^{\prime }J_{\ell }}-\tilde{\nu}_{1}\right) +\dfrac{1}{2}\Phi
_{ba}^{\ast }\left( \nu _{J_{u}J_{\ell }}-\tilde{\nu}_{1}\right) \right] %
\left[ \dfrac{1}{2}\Phi _{ba}\left( \nu _{J_{u}^{\prime }J_{\ell }^{\prime
}}-\tilde{\nu}\right) +\dfrac{1}{2}\Phi _{ba}^{\ast }\left( \nu
_{J_{u}J_{\ell }^{\prime }}-\tilde{\nu}\right) \right] \right\}%
\end{array}%
\ .  \label{eq -- redist fine zero B}
\end{equation}

The second branching ratio is made of two subtracted terms. In the first
term, $\Gamma _{R}+\Gamma _{I}+\tfrac{1}{2}\left[ D^{(K)}(\alpha
_{u}J_{u})+D^{(K)}(\alpha _{u}J_{u}^{\prime })\right] $ is the inverse
lifetime of the upper level population or coherence (alignment), which is
associated with a corresponding level width by virtue of the Heisenberg
uncertainty principle. In the second term and also in the first branching
ratio, $\Gamma _{R}+\Gamma _{I}+\Gamma _{E}$ is the line width %
\citep{Baranger-58}. In the presence of elastic or quasi-elastic collisions, 
$\Gamma _{E}$ and $\tfrac{1}{2}\left[ D^{(K)}(\alpha
_{u}J_{u})+D^{(K)}(\alpha _{u}J_{u}^{\prime })\right] $ are both non-zero
and are different from each other. In other words, when there are elastic or
quasi-elastic collisions, the level width and line width are different. %
\citet{Sahal-Bommier-17} develop Eq. (61) in \citet{Baranger-58} and show
that interference terms between the upper and lower level may contribute to
the line width without contributing to each level width. It is the same with
the purely elastic effects, in which the atom does not change Zeeman
sublevel, and which contribute to $\Gamma _{E}$ but not to $\tfrac{1}{2}%
\left[ D^{(K)}(\alpha _{u}J_{u})+D^{(K)}(\alpha _{u}J_{u}^{\prime })\right] $%
.

The branching ratio in line 6 of Eq. (\ref{eq -- redist fine zero B}) can be
rewritten as%
\begin{equation}
\begin{array}{l}
\medskip \dfrac{\Gamma _{R}}{\Gamma _{R}+\Gamma _{I}+\dfrac{1}{2}\left[
D^{(K)}(\alpha _{u}J_{u})+D^{(K)}(\alpha _{u}J_{u}^{\prime })\right] +\dfrac{%
\mathrm{i}\Delta E_{J_{u}J_{u}^{\prime }}}{\hbar }}-\dfrac{\Gamma _{R}}{%
\Gamma _{R}+\Gamma _{I}+\Gamma _{E}+\dfrac{\mathrm{i}\Delta
E_{J_{u}J_{u}^{\prime }}}{\hbar }} \\ 
\medskip =\dfrac{\Gamma _{R}}{\Gamma _{R}+\Gamma _{I}+\dfrac{1}{2}\left[
D^{(K)}(\alpha _{u}J_{u})+D^{(K)}(\alpha _{u}J_{u}^{\prime })\right] +\dfrac{%
\mathrm{i}\Delta E_{J_{u}J_{u}^{\prime }}}{\hbar }}\times \dfrac{\Gamma _{E}-%
\dfrac{1}{2}\left[ D^{(K)}(\alpha _{u}J_{u})+D^{(K)}(\alpha
_{u}J_{u}^{\prime })\right] }{\Gamma _{R}+\Gamma _{I}+\Gamma _{E}+\dfrac{%
\mathrm{i}\Delta E_{J_{u}J_{u}^{\prime }}}{\hbar }}%
\end{array}%
\ .
\end{equation}%
\citet{Belluzzi-TrujilloB-14}, who totally ignore the depolarizing collision
contribution via the $D^{(K)}$ coefficients, have instead (see their Eqs.
(26) and (36))%
\begin{equation}
\dfrac{\Gamma _{R}}{\Gamma _{R}+\Gamma _{I}+\dfrac{\mathrm{i}\Delta
E_{J_{u}J_{u}^{\prime }}}{\hbar }}\times \dfrac{\Gamma _{E}}{\Gamma
_{R}+\Gamma _{I}+\Gamma _{E}}\ ,
\end{equation}%
which differs in the $\Delta E_{J_{u}J_{u}^{\prime }}$ contribution, but
coincides with ours when $\Delta E_{J_{u}J_{u}^{\prime }}=0$. As for the
other branching ratio given in the first term of line 5 of Eq. (\ref{eq --
redist fine zero B}), they have for this branching ratio (see their Eqs.
(30) and (36))%
\begin{equation}
\dfrac{\Gamma _{R}}{\Gamma _{R}+\Gamma _{I}+\dfrac{\mathrm{i}\Delta
E_{J_{u}J_{u}^{\prime }}}{\hbar }}\times \dfrac{\Gamma _{R}+\Gamma _{I}}{%
\Gamma _{R}+\Gamma _{I}+\Gamma _{E}}
\end{equation}%
while we use a different expression:%
\begin{equation}
\dfrac{\Gamma _{R}}{\Gamma _{R}+\Gamma _{I}+\Gamma _{E}+\dfrac{\mathrm{i}%
\Delta E_{J_{u}J_{u}^{\prime }}}{\hbar }}\ .
\end{equation}%
Again, their branching ratio coincides with ours when $\Delta
E_{J_{u}J_{u}^{\prime }}=0$. As for the product profile given in line 7 of
Eq. (\ref{eq -- redist fine zero B}), they have instead%
\begin{equation}
\left[ \dfrac{1}{2}\Phi _{ba}\left( \nu _{J_{u}J_{\ell }}-\tilde{\nu}%
_{1}\right) +\dfrac{1}{2}\Phi _{ba}^{\ast }\left( \nu _{J_{u}J_{\ell }}-%
\tilde{\nu}_{1}\right) \right] \left[ \dfrac{1}{2}\Phi _{ba}\left( \nu
_{J_{u}^{\prime }J_{\ell }^{\prime }}-\tilde{\nu}\right) +\dfrac{1}{2}\Phi
_{ba}^{\ast }\left( \nu _{J_{u}J_{\ell }^{\prime }}-\tilde{\nu}\right) %
\right] \ ,
\end{equation}%
where $J_{u}$ instead of $J_{u}^{\prime }$ is in the first profile of the
formula. In addition, we would obtain a formula similar to their Eq. (26)
for the Racah coefficients only in the case of an isolated $J_{\ell }$ level.

In the collisionless regime, this redistribution function with the branching
ratio (the coefficient before the $\delta $\ function) is in agreement with
that derived in the metalevel heuristic approach by \citet{Landi-etal-97},
and also derived by \citet{Casini-etal-14}.

\subsection{Considering also hyperfine structure}

In the case of non-zero hyperfine structure and zero magnetic field, we find%
\begin{equation}
\begin{array}{l}
\medskip \mathcal{R}_{ij}\left( \nu ,\nu _{1},\vec{\Omega},\vec{\Omega}_{1};%
\vec{B}=\vec{0}\right) =\dsum\limits_{J_{u}F_{u}J_{u}^{\prime }F_{u}^{\prime
}J_{\ell }F_{\ell }J_{\ell }^{\prime }F_{\ell }^{\prime }KQ}\int f(\vec{v})%
\mathrm{d}^{3}\vec{v}\ (-1)^{Q}\mathcal{T}_{-Q}^{K}(j,\vec{\Omega}_{1})%
\mathcal{T}_{Q}^{K}(i,\vec{\Omega}) \\ 
\medskip \times 3\dfrac{2L_{u}+1}{2S+1}(2J_{u}+1)(2J_{u}^{\prime
}+1)(2J_{\ell }+1)(2J_{\ell }^{\prime }+1)(2F_{u}+1)(2F_{u}^{\prime
}+1)(2F_{\ell }+1)(2F_{\ell }^{\prime }+1)\ (-1)^{F_{\ell }-F_{\ell
}^{\prime }} \\ 
\medskip \times \left\{ 
\begin{array}{ccc}
J_{u} & 1 & J_{\ell } \\ 
L_{\ell } & S & L_{u}%
\end{array}%
\right\} \left\{ 
\begin{array}{ccc}
J_{u}^{\prime } & 1 & J_{\ell } \\ 
L_{\ell } & S & L_{u}%
\end{array}%
\right\} \left\{ 
\begin{array}{ccc}
J_{u} & 1 & J_{\ell }^{\prime } \\ 
L_{\ell } & S & L_{u}%
\end{array}%
\right\} \left\{ 
\begin{array}{ccc}
J_{u}^{\prime } & 1 & J_{\ell }^{\prime } \\ 
L_{\ell } & S & L_{u}%
\end{array}%
\right\} \\ 
\medskip \times \left\{ 
\begin{array}{ccc}
F_{u} & 1 & F_{\ell } \\ 
J_{\ell } & I & J_{u}%
\end{array}%
\right\} \left\{ 
\begin{array}{ccc}
F_{u}^{\prime } & 1 & F_{\ell } \\ 
J_{\ell } & I & J_{u}^{\prime }%
\end{array}%
\right\} \left\{ 
\begin{array}{ccc}
F_{u} & 1 & F_{\ell }^{\prime } \\ 
J_{\ell } & I & J_{u}%
\end{array}%
\right\} \left\{ 
\begin{array}{ccc}
F_{u}^{\prime } & 1 & F_{\ell }^{\prime } \\ 
J_{\ell } & I & J_{u}^{\prime }%
\end{array}%
\right\} \\ 
\medskip \times \left\{ 
\begin{array}{ccc}
K & F_{u} & F_{u}^{\prime } \\ 
F_{\ell } & 1 & 1%
\end{array}%
\right\} \left\{ 
\begin{array}{ccc}
K & F_{u} & F_{u}^{\prime } \\ 
F_{\ell }^{\prime } & 1 & 1%
\end{array}%
\right\} \\ 
\medskip \times \left\{ \dfrac{\Gamma _{R}}{\Gamma _{R}+\Gamma _{I}+\Gamma
_{E}+\dfrac{\mathrm{i}\Delta E_{F_{u}F_{u}^{\prime }}}{\hbar }}\ \delta
\left( \tilde{\nu}-\tilde{\nu}_{1}-\nu _{F_{\ell }F_{\ell }^{\prime
}}\right) \left[ \dfrac{1}{2}\Phi _{ba}\left( \nu _{F_{u}^{\prime }F_{\ell
}}-\tilde{\nu}_{1}\right) +\dfrac{1}{2}\Phi _{ba}^{\ast }\left( \nu
_{F_{u}F_{\ell }}-\tilde{\nu}_{1}\right) \right] \right. \\ 
\medskip +\left[ \dfrac{\Gamma _{R}}{\Gamma _{R}+\Gamma _{I}+\dfrac{1}{2}%
\left[ D^{(K)}(\alpha _{u}F_{u})+D^{(K)}(\alpha _{u}F_{u}^{\prime })\right] +%
\dfrac{\mathrm{i}\Delta E_{F_{u}F_{u}^{\prime }}}{\hbar }}-\dfrac{\Gamma _{R}%
}{\Gamma _{R}+\Gamma _{I}+\Gamma _{E}+\dfrac{\mathrm{i}\Delta
E_{F_{u}F_{u}^{\prime }}}{\hbar }}\right] \\ 
\medskip \times \left. \left[ \dfrac{1}{2}\Phi _{ba}\left( \nu
_{F_{u}^{\prime }F_{\ell }}-\tilde{\nu}_{1}\right) +\dfrac{1}{2}\Phi
_{ba}^{\ast }\left( \nu _{F_{u}F_{\ell }}-\tilde{\nu}_{1}\right) \right] %
\left[ \dfrac{1}{2}\Phi _{ba}\left( \nu _{F_{u}^{\prime }F_{\ell }^{\prime
}}-\tilde{\nu}\right) +\dfrac{1}{2}\Phi _{ba}^{\ast }\left( \nu
_{F_{u}F_{\ell }^{\prime }}-\tilde{\nu}\right) \right] \right\}%
\end{array}%
\ ,
\end{equation}%
where in the energy difference $\Delta E_{F_{u}F_{u}^{\prime }}$ each $F_{u}$
quantum number stands for all the quantum numbers $\left( \alpha
L_{u}SJ_{u}IF_{u}\right) $ defining the $F_{u}$ level. As for the
depolarizing collisional rate $D^{(K)}(\alpha _{u}F_{u})$, $\alpha _{u}$
stands for $\left( \alpha L_{u}SJ_{u}I\right) $.

As stated at the end of Sect. \ref{section -- stateq}, in this case it was
possible to introduce the contribution of the depolarizing collisions within
each fine or hyperfine structure level, via the $D^{(K)}$ coefficients. This
was possible because in zero magnetic fields the energy differences $\Delta
E $ in the branching ratios no longer depend on the Zeeman quantum numbers $M
$, which implies that the summations over these indices can be performed.
However, as stated below in greater detail, this does not completely account
for the depolarizing effect of the collisions with neutral hydrogen atoms,
because the depolarization is also due to the collisional transitions
between the fine or hyperfine structure levels themselves, which cannot
easily be taken into account in the present formalism. This requires the
numerical resolution of the statistical equilibrium.

In the collisionless regime, this redistribution function with the branching
ratio (the coefficient before the $\delta $\ function) is in agreement with
that derived in the metalevel heuristic approach by \citet{Landi-etal-97},
and also derived by \citet{Casini-etal-14}.

\section{Conclusion}

\label{section -- conclusion}

In the present paper, we have given expressions for the radiation
redistribution function describing both coherent and incoherent scattering
in the presence of magnetic field, fine and hyperfine structure. It is found
that the magnetic field (the Zeeman splitting) and the fine or hyperfine
structure affect the form of the branching ratios, which weight both
coherent and incoherent scattering contributions to the redistribution
function. Thus, the redistribution function cannot be factorized in terms of
the different contributions from individual redistributions in frequency and
directions because all mechanisms -- magnetic fields, fine and hyperfine
structures, line profiles, velocities, and Doppler effect, spherical tensors
for polarimetry -- are intermingled in the redistribution function.

Since the pioneering work by \citet{Omont-etal-72}, it is well-known that
the balance between the frequency coherent and incoherent contributions is
dominated by the elastic collisions, which contribute to the line
broadening. In a classical representation, they scramble the phase of the
radiating oscillator, thus introducing finite coherence time intervals,
which correspond to non-zero energy intervals via the Heisenberg uncertainty
principle. But a quantitative modeling of the radiation emitted or scattered
by an atom requires a quantum description. These collisions then become
responsible for transitions between Zeeman sublevels. When there is no
magnetic field, and all the Zeeman sublevels of a level are isoenergetic,
these transitions are strictly elastic. When there is a magnetic field,
these transitions are not exactly elastic. However, as the Zeeman splitting
is typically very small with respect to the energy difference between the
two levels connected by the emitted line, these transitions may be
considered quasi-elastic.\ There are also quasi-elastic transitions between
the fine and hyperfine structure levels of\ a given term. In addition, in
the line broadening, there is also a contribution of collisions that do not
even cause transitions. There are also interference terms that contribute to
the line broadening, as noted by \citet{Sahal-Bommier-14,Sahal-Bommier-17}.
The line broadening effect is fully accounted for via the $\Gamma _{E}$ term
in the above redistribution function expressions.

In the solar atmosphere, these collisions are essentially between the
emitting atom or ion and neutral hydrogen atoms, as described in Sect. 4 of %
\citet{Bommier-16a}.\ Collisions that cause transitions between levels or
sublevels are responsible for depolarization. This is easily understood in
the case of transitions between Zeeman sublevels of a given level. The
observed linear polarization by scattering is due to imbalance between
Zeeman sublevel populations. These collisions tend to equalize the sublevel
populations, thus reducing the linear polarization in the scattered
radiation.\ They are accounted for via the $D^{(K)}$ term (with $D^{(0)}=0$%
), which can be easily introduced in the above expressions only in the
absence of a magnetic field. In the presence of a magnetic field the
depolarizing effect of the transitions between Zeeman sublevels cannot be
simply accounted for by this term, and there are also transitions between
fine and hyperfine structure levels of a given level or term, and these
transitions also contribute to depolarization.\ They cannot be taken into
account at all in the above expressions.

All the corresponding rates, namely, transfer rates between fine or
hyperfine structure levels, depolarizing rates and line broadening, are of
the same order of magnitude. An example is given in the case of the %
\ion{Na}{i} $D_{1}$-$D_{2}$ lines by \citet{Kerkeni-Bommier-02}. Numerical
values of the corresponding $g$ coefficients as functions of temperature and
neutral hydrogen density are given in Eqs. (11-16) of that paper. It can be
seen that the $g^{1}$ factors, which enter the line broadening, are of the
same order of magnitude as the $g^{2}$ factors that are responsible for the
depolarization due to transitions between Zeeman sublevels, and are also of
the same order of magnitude as the $g^{0}$ factors that are responsible for
the transfer rates of the collisional transitions between the fine and
hyperfine structure levels.

The transitions between the fine and hyperfine structure levels play a
depolarizing role as important as the one of the transitions between Zeeman
sublevels, but they can be accounted for only in a numerical solution of the
statistical equilibrium. As an example, we numerically solved the
statistical equilibrium under solar upper photosphere conditions. In
90-degree scattering, the \ion{Na}{i} $D_{2}$ linear polarization rate would
be $5.5\times 10^{-3}$ in the absence of collisions. When all the
collisional transitions are accounted for, between the Zeeman sublevels and
between the fine structure levels $3p^{2}P_{1/2}$ and $3p^{2}P_{3/2}$, the
scattered light polarization rate decreases to $2.2\times 10^{-3}$. When the
transition rates between the fine structure levels $3p^{2}P_{1/2}$ and $%
3p^{2}P_{3/2}$ are set to zero, the scattered light polarization rate
increases to $3.6\times 10^{-3}$. It can be concluded that the transitions
between fine structure levels are as important as the transitions between
Zeeman sublevels for depolarizing the emitted line.

Fully taking into account all these collisional transitions requires a
numerical solution of the statistical equilibrium of the atomic levels and
coherences.\ This is not the case for the redistribution function given
above, which is based on an analytical solution of the statistical
equilibrium, where in particular the transfer rates between the fine and
hyperfine structure levels cannot be taken into account.\ The above
redistribution functions thus cannot completely account for the depolarizing
effect of collisions. However, such functions are widely used in radiative
transfer modeling of stellar atmospheres. They ignore the lower level
alignment, which also requires the numerical solution of the statistical
equilibrium system of equations. In the present paper, the redistribution
function derivation has been made under the unpolarized and infinitely sharp
lower level approximation. An innovative numerical approach alternatively
based on the numerical solution of the statistical equilibrium is currently
in progress \citep{Bommier-16b}, which enables a complete account of the
collisional effects. This approach does not make use of a redistribution
function.

\begin{acknowledgements}
The author is greatly indebted to the referee of this paper, Roberto Casini, for helpful suggestions that improved the quality of the manuscript.
\end{acknowledgements}

\appendix

\section{Redistribution matrix in incomplete Paschen-Back effect}

\label{Appendix}

In this Appendix, we provide the redistribution matrix in the case of
incomplete Paschen-Back effect, as introduced in Sect. \ref{subsect --
Paschen-Back}.

\subsection{Considering only fine structure}

The redistribution function is given by

\begin{equation}
\begin{array}{l}
\medskip \mathcal{R}_{ij}\left( \nu ,\nu _{1},\vec{\Omega},\vec{\Omega}_{1};%
\vec{B}\right) =\dsum\limits_{J_{u}\bar{J}_{u}J_{u}^{\ast
}M_{u}J_{u}^{\prime }\bar{J}_{u}^{\prime }J_{u}^{\prime \ast }M_{u}^{\prime
}J_{\ell }\bar{J}_{\ell }J_{\ell }^{\ast }M_{\ell }J_{\ell }^{\prime }\bar{J}%
_{\ell }^{\prime }J_{\ell }^{\prime \ast }M_{\ell }^{\prime }KK^{\prime
}Q}\int f(\vec{v})\mathrm{d}^{3}\vec{v}\ (-1)^{Q}\mathcal{T}_{-Q}^{K^{\prime
}}(j,\vec{\Omega}_{1})\mathcal{T}_{Q}^{K}(i,\vec{\Omega}) \\ 
\medskip \times 3\dfrac{2L_{u}+1}{2S+1}\sqrt{(2K+1)(2K^{\prime }+1)}\
(-1)^{M_{\ell }-M_{\ell }^{\prime }} \\ 
\medskip \times \sqrt{(2J_{u}+1)(2\bar{J}_{u}+1)(2J_{u}^{\prime }+1)(2\bar{J}%
_{u}^{\prime }+1)(2J_{\ell }+1)(2\bar{J}_{\ell }+1)(2J_{\ell }^{\prime }+1)(2%
\bar{J}_{\ell }^{\prime }+1)} \\ 
\medskip \times C_{J_{u}^{\ast }M_{u}}^{J_{u}}\left( B\right) C_{J_{u}^{\ast
}M_{u}}^{\bar{J}_{u}}\left( B\right) C_{J_{u}^{\prime \ast }M_{u}^{\prime
}}^{J_{u}^{\prime }}\left( B\right) C_{J_{u}^{\prime \ast }M_{u}^{\prime }}^{%
\bar{J}_{u}^{\prime }}\left( B\right) C_{J_{\ell }^{\ast }M_{\ell
}}^{J_{\ell }}\left( B\right) C_{J_{\ell }^{\ast }M_{\ell }}^{\bar{J}_{\ell
}}\left( B\right) C_{J_{\ell }^{\prime \ast }M_{\ell }^{\prime }}^{J_{\ell
}^{\prime }}\left( B\right) C_{J_{\ell }^{\prime \ast }M_{\ell }^{\prime }}^{%
\bar{J}_{\ell }^{\prime }}\left( B\right) \\ 
\medskip \times \left\{ 
\begin{array}{ccc}
J_{u} & 1 & J_{\ell } \\ 
L_{\ell } & S & L_{u}%
\end{array}%
\right\} \left\{ 
\begin{array}{ccc}
J_{u}^{\prime } & 1 & \bar{J}_{\ell } \\ 
L_{\ell } & S & L_{u}%
\end{array}%
\right\} \left\{ 
\begin{array}{ccc}
\bar{J}_{u} & 1 & J_{\ell }^{\prime } \\ 
L_{\ell } & S & L_{u}%
\end{array}%
\right\} \left\{ 
\begin{array}{ccc}
\bar{J}_{u}^{\prime } & 1 & \bar{J}_{\ell }^{\prime } \\ 
L_{\ell } & S & L_{u}%
\end{array}%
\right\} \\ 
\medskip \times \left( 
\begin{array}{ccc}
J_{u} & 1 & J_{\ell } \\ 
-M_{u} & p & M_{\ell }%
\end{array}%
\right) \left( 
\begin{array}{ccc}
J_{u}^{\prime } & 1 & \bar{J}_{\ell } \\ 
-M_{u}^{\prime } & p^{\prime } & M_{\ell }%
\end{array}%
\right) \left( 
\begin{array}{ccc}
\bar{J}_{u} & 1 & J_{\ell }^{\prime } \\ 
-M_{u} & p^{\prime \prime \prime } & M_{\ell }^{\prime }%
\end{array}%
\right) \left( 
\begin{array}{ccc}
\bar{J}_{u}^{\prime } & 1 & \bar{J}_{\ell }^{\prime } \\ 
-M_{u}^{\prime } & p^{\prime \prime } & M_{\ell }^{\prime }%
\end{array}%
\right) \\ 
\medskip \times \left( 
\begin{array}{ccc}
1 & 1 & K^{\prime } \\ 
-p & p^{\prime } & Q%
\end{array}%
\right) \left( 
\begin{array}{ccc}
1 & 1 & K \\ 
-p^{\prime \prime \prime } & p^{\prime \prime } & Q%
\end{array}%
\right) \\ 
\medskip \times \left\{ \dfrac{\Gamma _{R}}{\Gamma _{R}+\Gamma _{I}+\Gamma
_{E}+\dfrac{\mathrm{i}\Delta E_{M_{u}M_{u}^{\prime }}}{\hbar }}\ \delta
\left( \tilde{\nu}-\tilde{\nu}_{1}-\nu _{M_{\ell }M_{\ell }^{\prime
}}\right) \left[ \dfrac{1}{2}\Phi _{ba}\left( \nu _{M_{u}^{\prime }M_{\ell
}}-\tilde{\nu}_{1}\right) +\dfrac{1}{2}\Phi _{ba}^{\ast }\left( \nu
_{M_{u}M_{\ell }}-\tilde{\nu}_{1}\right) \right] \right. \\ 
\medskip +\left[ \dfrac{\Gamma _{R}}{\Gamma _{R}+\Gamma _{I}+\dfrac{\mathrm{i%
}\Delta E_{M_{u}M_{u}^{\prime }}}{\hbar }}-\dfrac{\Gamma _{R}}{\Gamma
_{R}+\Gamma _{I}+\Gamma _{E}+\dfrac{\mathrm{i}\Delta E_{M_{u}M_{u}^{\prime }}%
}{\hbar }}\right] \\ 
\medskip \times \left. \left[ \dfrac{1}{2}\Phi _{ba}\left( \nu
_{M_{u}^{\prime }M_{\ell }}-\tilde{\nu}_{1}\right) +\dfrac{1}{2}\Phi
_{ba}^{\ast }\left( \nu _{M_{u}M_{\ell }}-\tilde{\nu}_{1}\right) \right] %
\left[ \dfrac{1}{2}\Phi _{ba}\left( \nu _{M_{u}^{\prime }M_{\ell }^{\prime
}}-\tilde{\nu}\right) +\dfrac{1}{2}\Phi _{ba}^{\ast }\left( \nu
_{M_{u}M_{\ell }^{\prime }}-\tilde{\nu}\right) \right] \right\}%
\end{array}%
\ .
\end{equation}

\subsection{Considering also hyperfine structure}

Given the quantum numbers in zero magnetic field $\left( JIFM\right) $, it
is possible to obtain the unique $M_{J}$ quantum number of this level, from
the relation%
\begin{equation}
M_{J}=\min \left( M+I,J\right) -\left( J+I-F\right) \ ,
\label{eq -- findmj>0}
\end{equation}%
where $\min \left( M+I,J\right) $ is the smallest algebraic of the two
quantities $M+I$ and $J$, and where the minimum is taken in algebraic value
taking into account the sign of $M$. This equation is valid if the Land\'{e}
factor $g_{J}$ is positive and when the level hyperfine structure has
increasing $F$\ values with increasing energies. When the hyperfine
structure is reversed, i.e. when when the level hyperfine structure has
decreasing $F$\ values with increasing energies, always with positive $g_{J}$%
, the formula is instead%
\begin{equation}
M_{J}=-\min \left( -M+I,J\right) +\left( J+I-F\right) \ .
\label{eq -- findmj<0}
\end{equation}%
Instead, when the Land\'{e} factor $g_{J}$ is negative, Eq. (\ref{eq --
findmj<0}) applies to normal hyperfine structure and Eq. (\ref{eq --
findmj>0}) applies $to$ reversed hyperfine structure. Conversely, given $%
\left( JIM_{J}M\right) $\ in zero magnetic field, $F$ can be retrieved
following%
\begin{equation}
F=\left( J+I\right) -\left( \min \left( M+I,J\right) -M_{J}\right)
\label{eq -- findf>0}
\end{equation}%
for normal hyperfine structure, and%
\begin{equation}
F=\left( J+I\right) -\left( \min \left( -M+I,J\right) +M_{J}\right)
\label{eq -- findf<0}
\end{equation}%
for reverse hyperfine structure, both for positive Land\'{e} factor $g_{J}$.
For negative Land\'{e} factor $g_{J}$, Eq. (\ref{eq -- findf<0}) applies to
normal hyperfine structure and Eq. (\ref{eq -- findf>0}) applies to reversed
hyperfine structure.

Given these equations, it is possible to determine what decoupling weights $%
C_{F^{\ast }M}^{F^{\prime }}\left( B\right) $ have to be applied, even if
the fine structure $J$\ quantum number is $J^{\ast }$\ and no longer a good
quantum number. The procedure is then as follows. Given $\left( J^{\ast
}IF^{\ast }M\right) $, it is possible to determine the corresponding $M_{J}$%
\ value as described above. Then, the weights $C_{J^{\ast }M_{J}}^{J^{\prime
}}\left( B\right) $ which express the partial decoupling of the fine
structure in the presence of the magnetic field, can be computed. We then
denote these weights $C_{J^{\ast }M_{J}\left( F^{\ast }M\right) }^{J^{\prime
}}\left( B\right) $,\ in order to keep in mind that we started from $\left(
J^{\ast }IF^{\ast }M\right) $. This coefficient introduces the good quantum
number $J^{\prime }$, associated with the same value of the magnetic quantum
number $M_{J}$. Then, always using the above formul\ae , a new corresponding 
$F$\ value can be determined, which is denoted as $F^{\ast \ast }$ because
it is not a good quantum number, and it is different from $F^{\ast }$\ when $%
J^{\prime }$\ is different from $J^{\ast }$. Then, the corresponding $%
C_{F^{\ast \ast }M}^{F^{\prime }}\left( B\right) $ weights for the
development over the $F^{\prime }$\ good quantum numbers can be determined,
which is denoted $C_{F^{\ast \ast }\left( J^{\prime }M_{J}\right)
M}^{F^{\prime }}\left( B\right) $ to keep in mind that $F^{\ast \ast }$ was
determined from $\left( J^{\prime }IM_{J}M\right) $.

The redistribution function is given by

\begin{equation}
\begin{array}{l}
\medskip \mathcal{R}_{ij}\left( \nu ,\nu _{1},\vec{\Omega},\vec{\Omega}_{1};%
\vec{B}\right) =\dsum\limits_{J_{u}\bar{J}_{u}J_{u}^{\ast }F_{u}\bar{F}%
_{u}F_{u}^{\ast }M_{u}J_{u}^{\prime }\bar{J}_{u}^{\prime }J_{u}^{\prime \ast
}F_{u}^{\prime }\bar{F}_{u}^{\prime }F_{u}^{\prime \ast }M_{u}^{\prime
}J_{\ell }\bar{J}_{\ell }J_{\ell }^{\ast }F_{\ell }\bar{F}_{\ell }F_{\ell
}^{\ast }M_{\ell }J_{\ell }^{\prime }\bar{J}_{\ell }^{\prime }J_{\ell
}^{\prime \ast }F_{\ell }^{\prime }\bar{F}_{\ell }^{\prime }F_{\ell
}^{\prime \ast }M_{\ell }^{\prime }KK^{\prime }Q} \\ 
\medskip \times \int f(\vec{v})\mathrm{d}^{3}\vec{v}\ (-1)^{Q}\mathcal{T}%
_{-Q}^{K^{\prime }}(j,\vec{\Omega}_{1})\mathcal{T}_{Q}^{K}(i,\vec{\Omega})
\\ 
\medskip \times 3\dfrac{2L_{u}+1}{2S+1}\sqrt{(2K+1)(2K^{\prime }+1)}\
(-1)^{M_{\ell }-M_{\ell }^{\prime }} \\ 
\medskip \times \sqrt{(2J_{u}+1)(2\bar{J}_{u}+1)(2J_{u}^{\prime }+1)(2\bar{J}%
_{u}^{\prime }+1)(2J_{\ell }+1)(2\bar{J}_{\ell }+1)(2J_{\ell }^{\prime }+1)(2%
\bar{J}_{\ell }^{\prime }+1)} \\ 
\medskip \times \sqrt{(2F_{u}+1)(2\bar{F}_{u}+1)(2F_{u}^{\prime }+1)(2\bar{F}%
_{u}^{\prime }+1)(2F_{\ell }+1)(2\bar{F}_{\ell }+1)(2F_{\ell }^{\prime }+1)(2%
\bar{F}_{\ell }^{\prime }+1)} \\ 
\medskip \times C_{J_{u}^{\ast }M_{J_{u}}\left( F_{u}^{\ast }M_{u}\right)
}^{J_{u}}\left( B\right) C_{J_{u}^{\ast }M_{J_{u}}\left( F_{u}^{\ast
}M_{u}\right) }^{\bar{J}_{u}}\left( B\right) C_{J_{u}^{\prime \ast
}M_{J_{u}}^{\prime }\left( F_{u}^{\prime \ast }M_{u}^{\prime }\right)
}^{J_{u}^{\prime }}\left( B\right) C_{J_{u}^{\prime \ast }M_{J_{u}}^{\prime
}\left( F_{u}^{\prime \ast }M_{u}^{\prime }\right) }^{\bar{J}_{u}^{\prime
}}\left( B\right) \\ 
\medskip \times C_{J_{\ell }^{\ast }M_{J_{\ell }}\left( F_{\ell }^{\ast
}M_{\ell }\right) }^{J_{\ell }}\left( B\right) C_{J_{\ell }^{\ast
}M_{J_{\ell }}\left( F_{\ell }^{\ast }M_{\ell }\right) }^{\bar{J}_{\ell
}}\left( B\right) C_{J_{\ell }^{\prime \ast }M_{J_{\ell }}^{\prime }\left(
F_{\ell }^{\prime \ast }M_{\ell }^{\prime }\right) }^{J_{\ell }^{\prime
}}\left( B\right) C_{J_{\ell }^{\prime \ast }M_{J_{\ell }}^{\prime }\left(
F_{\ell }^{\prime \ast }M_{\ell }^{\prime }\right) }^{\bar{J}_{\ell
}^{\prime }}\left( B\right) \\ 
\medskip \times C_{F_{u}^{\ast \ast }\left( J_{u}M_{J_{u}}\right)
M_{u}}^{F_{u}}\left( B\right) C_{\bar{F}_{u}^{\ast \ast }\left( \bar{J}%
_{u}M_{J_{u}}\right) M_{u}}^{\bar{F}_{u}}\left( B\right) C_{F_{u}^{\prime
\ast \ast }\left( J_{u}^{\prime }M_{J_{u}}^{\prime }\right) M_{u}^{\prime
}}^{F_{u}^{\prime }}\left( B\right) C_{\bar{F}_{u}^{\prime \ast \ast }\left( 
\bar{J}_{u}^{\prime }M_{J_{u}}^{\prime }\right) M_{u}^{\prime }}^{\bar{F}%
_{u}^{\prime }}\left( B\right) \\ 
\medskip \times C_{F_{\ell }^{\ast \ast }\left( J_{\ell }M_{J_{\ell
}}\right) M_{\ell }}^{F_{\ell }}\left( B\right) C_{\bar{F}_{\ell }^{\ast
\ast }\left( \bar{J}_{\ell }M_{J_{\ell }}\right) M_{\ell }}^{\bar{F}_{\ell
}}\left( B\right) C_{F_{\ell }^{\prime \ast \ast }\left( J_{\ell }^{\prime
}M_{J_{\ell }}^{\prime }\right) M_{\ell }^{\prime }}^{F_{\ell }^{\prime
}}\left( B\right) C_{\bar{F}_{\ell }^{\prime \ast \ast }\left( \bar{J}_{\ell
}^{\prime }M_{J_{\ell }}^{\prime }\right) M_{\ell }^{\prime }}^{\bar{F}%
_{\ell }^{\prime }}\left( B\right) \\ 
\medskip \times \left\{ 
\begin{array}{ccc}
J_{u} & 1 & J_{\ell } \\ 
L_{\ell } & S & L_{u}%
\end{array}%
\right\} \left\{ 
\begin{array}{ccc}
J_{u}^{\prime } & 1 & \bar{J}_{\ell } \\ 
L_{\ell } & S & L_{u}%
\end{array}%
\right\} \left\{ 
\begin{array}{ccc}
\bar{J}_{u} & 1 & J_{\ell }^{\prime } \\ 
L_{\ell } & S & L_{u}%
\end{array}%
\right\} \left\{ 
\begin{array}{ccc}
\bar{J}_{u}^{\prime } & 1 & \bar{J}_{\ell }^{\prime } \\ 
L_{\ell } & S & L_{u}%
\end{array}%
\right\} \\ 
\medskip \times \left\{ 
\begin{array}{ccc}
F_{u} & 1 & F_{\ell } \\ 
J_{\ell } & S & J_{u}%
\end{array}%
\right\} \left\{ 
\begin{array}{ccc}
F_{u}^{\prime } & 1 & \bar{F}_{\ell } \\ 
J_{\ell } & S & J_{u}%
\end{array}%
\right\} \left\{ 
\begin{array}{ccc}
\bar{F}_{u} & 1 & F_{\ell }^{\prime } \\ 
J_{\ell } & S & J_{u}%
\end{array}%
\right\} \left\{ 
\begin{array}{ccc}
\bar{F}_{u}^{\prime } & 1 & \bar{F}_{\ell }^{\prime } \\ 
J_{\ell } & S & J_{u}%
\end{array}%
\right\} \\ 
\medskip \times \left( 
\begin{array}{ccc}
F_{u} & 1 & F_{\ell } \\ 
-M_{u} & p & M_{\ell }%
\end{array}%
\right) \left( 
\begin{array}{ccc}
F_{u}^{\prime } & 1 & \bar{F}_{\ell } \\ 
-M_{u}^{\prime } & p^{\prime } & M_{\ell }%
\end{array}%
\right) \left( 
\begin{array}{ccc}
\bar{F}_{u} & 1 & F_{\ell }^{\prime } \\ 
-M_{u} & p^{\prime \prime \prime } & M_{\ell }^{\prime }%
\end{array}%
\right) \left( 
\begin{array}{ccc}
\bar{F}_{u}^{\prime } & 1 & \bar{F}_{\ell }^{\prime } \\ 
-M_{u}^{\prime } & p^{\prime \prime } & M_{\ell }^{\prime }%
\end{array}%
\right) \\ 
\medskip \times \left( 
\begin{array}{ccc}
1 & 1 & K^{\prime } \\ 
-p & p^{\prime } & Q%
\end{array}%
\right) \left( 
\begin{array}{ccc}
1 & 1 & K \\ 
-p^{\prime \prime \prime } & p^{\prime \prime } & Q%
\end{array}%
\right) \\ 
\medskip \times \left\{ \dfrac{\Gamma _{R}}{\Gamma _{R}+\Gamma _{I}+\Gamma
_{E}+\dfrac{\mathrm{i}\Delta E_{M_{u}M_{u}^{\prime }}}{\hbar }}\ \delta
\left( \tilde{\nu}-\tilde{\nu}_{1}-\nu _{M_{\ell }M_{\ell }^{\prime
}}\right) \left[ \dfrac{1}{2}\Phi _{ba}\left( \nu _{M_{u}^{\prime }M_{\ell
}}-\tilde{\nu}_{1}\right) +\dfrac{1}{2}\Phi _{ba}^{\ast }\left( \nu
_{M_{u}M_{\ell }}-\tilde{\nu}_{1}\right) \right] \right. \\ 
\medskip +\left[ \dfrac{\Gamma _{R}}{\Gamma _{R}+\Gamma _{I}+\dfrac{\mathrm{i%
}\Delta E_{M_{u}M_{u}^{\prime }}}{\hbar }}-\dfrac{\Gamma _{R}}{\Gamma
_{R}+\Gamma _{I}+\Gamma _{E}+\dfrac{\mathrm{i}\Delta E_{M_{u}M_{u}^{\prime }}%
}{\hbar }}\right] \\ 
\medskip \times \left. \left[ \dfrac{1}{2}\Phi _{ba}\left( \nu
_{M_{u}^{\prime }M_{\ell }}-\tilde{\nu}_{1}\right) +\dfrac{1}{2}\Phi
_{ba}^{\ast }\left( \nu _{M_{u}M_{\ell }}-\tilde{\nu}_{1}\right) \right] %
\left[ \dfrac{1}{2}\Phi _{ba}\left( \nu _{M_{u}^{\prime }M_{\ell }^{\prime
}}-\tilde{\nu}\right) +\dfrac{1}{2}\Phi _{ba}^{\ast }\left( \nu
_{M_{u}M_{\ell }^{\prime }}-\tilde{\nu}\right) \right] \right\}%
\end{array}%
\ .
\end{equation}

\end{document}